\documentclass[12pt]{article}
\usepackage{graphicx}

\def\hybrid{\topmargin 0pt      \oddsidemargin 0pt
        \headheight 0pt \headsep 0pt
       \voffset-1cm
        \textwidth 6.25in       
       \textheight 9.5in       
        \marginparwidth 0.0in
        \parskip 5pt plus 1pt   \jot = 1.5ex}
\catcode`\@=11
\def\marginnote#1{}

\newcount\hour
\newcount\minute
\newtoks\amorpm
\hour=\time\divide\hour by60
\minute=\time{\multiply\hour by60 \global\advance\minute by-\hour}
\edef\standardtime{{\ifnum\hour<12 \global\amorpm={am}%
        \else\global\amorpm={pm}\advance\hour by-12 \fi
        \ifnum\hour=0 \hour=12 \fi
        \number\hour:\ifnum\minute<10 0\fi\number\minute\the\amorpm}}
\edef\militarytime{\number\hour:\ifnum\minute<10 0\fi\number\minute}

\def\draftlabel#1{{\@bsphack\if@filesw {\let\thepage\relax
   \xdef\@gtempa{\write\@auxout{\string
      \newlabel{#1}{{\@currentlabel}{\thepage}}}}}\@gtempa
   \if@nobreak \ifvmode\nobreak\fi\fi\fi\@esphack}
        \gdef\@eqnlabel{#1}}
\def\@eqnlabel{}
\def\@vacuum{}
\def\draftmarginnote#1{\marginpar{\raggedright\scriptsize\tt#1}}

\def\draftlabel#1{{\@bsphack\if@filesw {\let\thepage\relax
   \xdef\@gtempa{\write\@auxout{\string
      \newlabel{#1}{{\@currentlabel}{\thepage}}}}}\@gtempa
   \if@nobreak \ifvmode\nobreak\fi\fi\fi\@esphack}
        \gdef\@eqnlabel{#1}}
\def\@eqnlabel{}
\def\@vacuum{}
\def\draftmarginnote#1{\marginpar{\raggedright\scriptsize\tt#1}}

\def\draft{\oddsidemargin -.5truein
        \def\@oddfoot{\sl preliminary draft \hfil
        \rm\thepage\hfil\sl\today\quad\militarytime}
        \let\@evenfoot\@oddfoot \overfullrule 3pt
        \let\label=\draftlabel
        \let\marginnote=\draftmarginnote
   \def\@eqnnum{(\theequation)\rlap{\kern\marginparsep\tt\@eqnlabel}%
\global\let\@eqnlabel\@vacuum}  }

\def\numberbysection{\@addtoreset{equation}{section}
        \def\theequation{\thesection.\arabic{equation}}}

\def\underline#1{\relax\ifmmode\@@underline#1\else
        $\@@underline{\hbox{#1}}$\relax\fi}

\def\titlepage{\@restonecolfalse\if@twocolumn\@restonecoltrue\onecolumn
     \else \newpage \fi \thispagestyle{empty}\c@page\z@
        \def\thefootnote{\fnsymbol{footnote}} }

\def\endtitlepage{\if@restonecol\twocolumn \else  \fi
        \def\thefootnote{\arabic{footnote}}
        \setcounter{footnote}{0}}  
\relax

\numberbysection
\hybrid

\newfont{\Bbb}{msbm10 scaled 1\@ptsize00}
\newfont{\Bbbb}{msbm7 scaled 1\@ptsize00}
\newcommand{\CC}{\mbox{\Bbb C}}

\newcommand{\DDD}{\raise-1pt\hbox{$\mbox{\Bbbb D}$}}

\newcommand{\UUU}{\raise-1pt\hbox{$\mbox{\Bbbb U}$}}

\newcommand{\ZZ}{\mbox{\Bbb Z}}
\newcommand{\z}{\raise-1pt\hbox{$\mbox{\Bbbb Z}$}}

\newcommand{\sss}{\raise-1pt\hbox{$\mbox{\Bbbb S}$}}

\def\beq{\begin{equation}}
\def\eeq{\end{equation}}

\newtheorem{lemma-definition}{Lemma-Definition}[section]

\def\normord{ {\scriptstyle {{\bullet}\atop{\bullet}}} }
\def\lbr{\left <}
\def\rbr{\right >}

\begin{document}

\begin{titlepage}

\title{Multi-component Pfaff-Toda hierarchy \\ within bilinear
formalism}

\author{A. Savchenko\thanks{
Skolkovo Institute of Science and Technology, 143026, Moscow, Russia
and National Research University Higher School of Economics,
20 Myasnitskaya Ulitsa,
Moscow 101000, Russia,
e-mail: ksanoobshhh@gmail.com}
\and
A.~Zabrodin\thanks{
National Research University Higher School of Economics,
20 Myasnitskaya Ulitsa,
Moscow 101000, Russia and
NRC ``Kurchatov institute'', Moscow, Russia;
e-mail: zabrodin@itep.ru}}

\date{November 2025}
\maketitle

\vspace{-7cm} \centerline{ \hfill ITEP-TH-36/25}\vspace{7cm}

\begin{abstract}

Using the free fermions technique and non-abelian 
bosonization rules 
we introduce the multi-component Pfaff-Toda 
hierarchy. The tau-function is defined as vacuum expectation
value of a Clifford group element of the algebra of Fermi-operators.
A generating bilinear integral equation for the 
tau-function is obtained.
A number of bilinear functional relations 
for the tau-function of the Hirota-Miwa type
are derived as corollaries of the generating bilinear equation.

\end{abstract}

\end{titlepage}

\vspace{5mm}

%

\tableofcontents

\vspace{5mm}

\section{Introduction}


The theory of integrable hierarchies of nonlinear differential 
and difference equations such as
Kadomtsev-Petviashvili (KP) and Toda lattice has a long and rich history.
It admits more than one seemingly different but equivalent formulations.
In the operator approach developed by the Kyoto school
\cite{DJKM83,JM83} the universal dependent variable of the hierarchies,
the tau-function, is represented as a vacuum expectation value of certain 
operators constructed from free fermions. 
For the Pfaff-type hierarchies (such as DKP also known as 
the coupled KP hierarchy and the Pfaff-Toda hierarchy,
see \cite{HO}--\cite{Takasaki09}),
these operators are exponential functions of quadratic forms
in the free charged fermions $\psi_n, \psi^*_n$ obeying the standard
anti-commutation relations:
\beq\label{int1}
g=\exp \left ( 
\sum_{j,k}\Bigl ( A_{jk}
\psi_{j}\psi^{*}_{k}+
B_{jk}
\psi_{j}\psi_{k}+
C_{jk}
\psi_{j}^{*}\psi^{*}_{k}\Bigr )
\right ).
\eeq
It is important to emphasize that
for the Pfaff-type hierarchies the quadratic forms in general
do not have a definite charge, i.e., in general
$B, C \neq 0$ (only the parity is definite),
whereas for the the KP or Toda hierarchies they are neutral,
i.e., $B=C=0$ for that cases. Operators of the form 
(\ref{int1}) are referred to as elements of the Clifford group.
They are known to obey the following bilinear operator identity:
\beq\label{int2}
\sum_{j\in \z}\Bigl (
\psi_{j}g \otimes
\psi_{j}^{*}g +
\psi_{j}^{*}g \otimes
\psi_{j}g \Bigr )
=\sum_{j\in \z}\Bigl (
g\psi_{j}\otimes g \psi_{j}^{*}+
g\psi_{j}^{*}\otimes g \psi_{j}\Bigr ).
\eeq
After sandwiching between certain 
states of the fermionic Fock space and 
bosonization it gives rise to bilinear equations for the 
tau-function, which is defined as vacuum expectation value of 
operators from the Clifford group. 

A natural generalization of this approach is to consider multi-component
fermions, i.e., assign an additional
index $\alpha \in \{1, \ldots , N\}$ (``a colour'') to the fermionic operators: 
$\psi_n^{(\alpha )}, \psi^{*(\alpha )}_n$
($N$-component charged fermions). Like in the one-component case,
the operators $\psi_n^{(\alpha )}$, $\psi^{*(\alpha )}_n$ carry charge $\pm 1$.
The corresponding theory is a source of matrix and 
multi-component hierarchies; for neutral quadratic forms one obtains in this way
multi-component KP and Toda hierarchies \cite{DJKM81}--\cite{TT07}.
In our paper \cite{SZ24}, we have applied the Fermi-operator approach to
introduce a natural multi-component extension of the DKP hierarchy,
and, among other things, basic bilinear relations 
for the tau-function have been
obtained. The starting point was the extension of the operator 
bilinear identity (\ref{int2}) to the $N$-component case,
\beq\label{int3}
\sum_{\gamma =1}^N \sum_{j\in \z}\Bigl (
\psi_{j}^{(\gamma )}g \otimes
\psi_{j}^{*(\gamma )}g +
\psi_{j}^{*(\gamma )}g \otimes
\psi_{j}^{(\gamma )}g \Bigr )
=\sum_{\gamma =1}^N \sum_{j\in \z}\Bigl (
g\psi_{j}^{(\gamma )}\otimes g \psi_{j}^{*(\gamma )}+
g\psi_{j}^{*(\gamma )}\otimes g \psi_{j}^{(\gamma )}\Bigr ),
\eeq
proved in \cite{SZ24}.
After sandwiching it between certain states from the Fock space and applying
non-abelian bosonization rules, the bilinear equations for the 
tau-function have been obtained in \cite{SZ24}. 

The aim of the present paper is to extend this approach to the more general case
of what can be named {\it multi-component Pfaff-Toda hierarchy}. The one-component
case was introduced and 
studied in detail by Takasaki in \cite{Takasaki09}.
In the $N$-component Pfaff-Toda hierarchy, the independent variables are
the ``times'',
\beq\label{times2}
\begin{array}{l}
{\bf t}=\{{\bf t}_1, {\bf t}_2, \ldots , {\bf t}_N\}, \qquad
{\bf t}_{\alpha}=\{t_{\alpha , 1}, t_{\alpha , 2}, t_{\alpha , 3}, 
\ldots \, \},
\\ \\
\bar {\bf t}=\{\bar {\bf t}_1, \bar {\bf t}_2, \ldots , 
\bar {\bf t}_N\}, \qquad
\bar {\bf t}_{\alpha}=\{\bar t_{\alpha , 1}, \bar t_{\alpha , 2}, 
\bar t_{\alpha , 3}, \ldots \, \},
\end{array}
\qquad \alpha = 1, \ldots , N
\eeq
(in general complex numbers) and two finite sets of integer variables
$$
{\bf n}=\{n_1, \ldots , n_N\}, \quad
\bar {\bf n}=\{\bar n_1, \ldots , \bar n_N\}, \quad n_{\alpha}, \bar n_{\alpha}
\in \ZZ .
$$
(Hereafter, the bar under any variable does not mean
complex conjugation, so the variables with and without bar should be regarded
as independent.)

We define the universal dependent variable (the tau-function) as
the vacuum expectation value
\beq\label{int4}
\tau ({\bf n}, \bar {\bf n}, {\bf t},
\bar {\bf t})=
\lbr {\bf n}\right | e^{J({\bf t})} g 
e^{-\bar J(\bar {\bf t})}\left |-\bar {\bf n}\rbr ,
\eeq
where $g$ is a Clifford group element of the general form (\ref{int1}).
The definition of
the left and right 
vacuum states $\lbr {\bf n}\right |$, $\left |-\bar {\bf n}\rbr$ and
the ``current operators'' $J({\bf t})$, $\bar J(\bar {\bf t})$ is standard.
It is given 
below in
section \ref{section:multi-component}.
The bilinear operator identity (\ref{int3}) implies that the tau-function
obeys an integral bilinear relation (see (\ref{main}) or (\ref{main1})
below) which
is referred to as the generating relation because contains, as its corollaries, 
the full set of bilinear relations with a finite number of terms  
(referred to as relations of the Hirota-Miwa type).

Note that if all the bar-variables are put equal 
to zero ($\bar {\bf n}={\bf 0}$,
$\bar {\bf t}={\bf 0}$), the tau-function (\ref{int4}), as a function of
${\bf n}, {\bf t}$, is a tau-function of the $N$-component DKP hierarchy
considered in \cite{SZ24} (the ``left'' copy of the DKP hierarchy). Similarly,
if one puts ${\bf n}={\bf 0}$, ${\bf t}={\bf 0}$, the tau-function,
as a function $\bar {\bf n}, \bar {\bf t}$, is a tau-function of 
another, ``right'', copy of the multi-component DKP hierarchy.

The structure of the paper is as follows. In section 2 
we introduce the notation and present necessary facts related to
the multicomponent fermions.
In section 3 we define the tau-function as expectation 
value of operators from the Clifford group and obtain the generating 
bilinear equation for it. Its general form 
is an integral functional relation for the 
tau-function. Section 4 is devoted to the derivation 
of bilinear equations of the Hirota-Miwa type from it.
We show that the minimal number of such terms is 4, and we obtain all
non-degenerate 4-point relations. In section 5, we discuss some perspectives
for further research.
The appendix contains the full list
of 2-point relations (still containing 4 terms) that follow from the 
4-point ones as a result of a degeneration procedure. We expect that these
relations are required for analysis of the dispersionless limit of
the multi-component Pfaff-Toda hierarchy (work in progress).

\section{Multi-component fermions}
\label{section:multi-component}

Following \cite{DJKM81,TT07}, 
we give here the necessary facts from the theory of multi-component free 
fermionic operators $\psi_{j}^{(\alpha )}$, 
$\psi_{j}^{*(\alpha )}$, where 
$j\in \ZZ$ is a discrete momentum index, and 
$\alpha =1, \ldots , N$ numbers different components.
These operators obey the anti-commutation relations
$$
[\psi_{j}^{(\alpha )}, \psi_{k}^{*(\beta )}]_+=\delta_{\alpha \beta}\delta_{jk},
\qquad
[\psi_{j}^{(\alpha )}, \psi_{k}^{(\beta )}]_+=
[\psi_{j}^{*(\alpha )}, \psi_{k}^{*(\beta )}]_+=0.
$$
We also introduce 
free fermionic fields constructed as series in the variable $z\in \CC$:
$$
\psi^{(\alpha )}(z)=\sum_{j\in \z}\psi^{(\alpha )}_j z^j,
\qquad
\psi^{*(\alpha )}(z)=\sum_{j\in \z}\psi^{*(\alpha )}_j z^{-j}.
$$

The Fock and dual Fock spaces are generated by action 
of creation operators to the vacuum states 
$\left | {\bf 0}\rbr$, $\lbr {\bf 0} \right |$ that satisfy the conditions
$$
\psi_{j}^{(\alpha )}\left | {\bf 0}\rbr =0 \quad (j<0), \qquad
\psi_{j}^{*(\alpha )}\left | {\bf 0}\rbr =0 \quad (j\geq 0),
$$
$$
\lbr {\bf 0}\right | \psi_{j}^{(\alpha )} =0 \quad (j\geq 0), \qquad
\lbr {\bf 0}\right | \psi_{j}^{*(\alpha )} =0 \quad (j< 0),
$$
so $\psi_{j}^{(\alpha )}$ with $j<0$ and $\psi_{j}^{*(\alpha )}$ with
$j\geq 0$ are annihilation operators while 
$\psi_{j}^{(\alpha )}$ with $j\geq 0$ and
$\psi_{j}^{*(\alpha )}$ with
$j<0$ are creation operators. 
Let ${\bf n}=\{ n_1, n_2, \ldots , n_N\}$ be a set
of $N$ integer numbers. We define the left and right vacuum states 
$\left | {\bf n}\rbr$, $\lbr {\bf n} \right |$
as
$$
\left | {\bf n}\rbr =\Psi_{n_N}^{*(N)}\ldots \Psi_{n_2}^{*(2)}
\Psi_{n_1}^{*(1)}\left | {\bf 0}\rbr , \qquad
\lbr {\bf n} \right |=\lbr {\bf 0} \right |\Psi_{n_1}^{(1)}\Psi_{n_2}^{(2)}\ldots
\Psi_{n_N}^{(N)},
$$
where
$$
\Psi_{n}^{*(\alpha )}=\left \{ \begin{array}{l}
\psi^{(\alpha )}_{n-1}\ldots \psi^{(\alpha )}_{0} \quad \,\,\, (n >0)
\\
1 \phantom{\psi^{(\alpha )}_{n-1}\ldots \psi^{(\alpha )}_{0}}
\quad (n=0)
\\
\psi^{*(\alpha )}_{n}\ldots \psi^{*(\alpha )}_{-1} \quad (n <0),
\end{array}
\right.
$$
$$
\Psi_{n}^{(\alpha )}=\left \{ \begin{array}{l}
\psi^{*(\alpha )}_{0}\ldots \psi^{*(\alpha )}_{n-1} \quad  \, (n >0)
\\
1 \phantom{\psi^{(\alpha )}_{n-1}\ldots \psi^{(\alpha )}_{0}}
\quad (n=0)
\\
\psi^{(\alpha )}_{-1}\ldots \psi^{(\alpha )}_{n} \quad \,\,\,\,\, (n <0).
\end{array}
\right.
$$

The modes of the current operators $J^{(\alpha )}(z)=
\normord \psi^{(\alpha )}(z)\psi^{*(\alpha )}(z)\normord$ have the form
$$
J_{k}^{(\alpha )}=\sum_{j\in \z}\normord 
\psi^{(\alpha )}_{j} \psi^{*(\alpha )}_{j+k}
\normord .
$$
Here the normal ordering $\normord (\ldots )\normord$ 
is defined by moving the annihilation operators 
to the right and creation operators to the left with 
the minus sign emerging each time
when two fermionic operators are permuted. Note that in the definition 
of $J_{k}^{(\alpha )}$ the normal ordering is essential only at
$k=0$. The commutation relations of these operators are
\beq\label{com1}
[J_k^{(\alpha )} , J_l^{(\beta )}]=k\delta_{\alpha \beta}\delta_{k, -l}.
\eeq

Let
\beq\label{times}
\begin{array}{l}
{\bf t}=\{{\bf t}_1, {\bf t}_2, \ldots , {\bf t}_N\}, \qquad
{\bf t}_{\alpha}=\{t_{\alpha , 1}, t_{\alpha , 2}, t_{\alpha , 3}, 
\ldots \, \},
\\ \\
\bar {\bf t}=\{\bar {\bf t}_1, \bar {\bf t}_2, \ldots , 
\bar {\bf t}_N\}, \qquad
\bar {\bf t}_{\alpha}=\{\bar t_{\alpha , 1}, \bar t_{\alpha , 2}, 
\bar t_{\alpha , 3}, \ldots \, \},
\end{array}
\qquad \alpha = 1, \ldots , N
\eeq
be $2N$ infinite sets of the independent time
variables (in general complex numbers). 
We introduce the operators
$$
J({\bf t})=\sum_{\alpha =1}^N \sum_{k\geq 1} t_{\alpha , k}J_k^{(\alpha )},
\qquad
\bar J(\bar {\bf t})=\sum_{\alpha =1}^N 
\sum_{k\geq 1} \bar t_{\alpha , k}J_{-k}^{(\alpha )}.
$$
Note the commutation relations
\beq\label{comm}
\begin{array}{l}
e^{J({\bf t})}\psi^{(\gamma )}(z)=e^{\xi ({\bf t}_{\gamma}, z)}
\psi^{(\gamma )}(z)e^{J({\bf t})}, \quad
e^{J({\bf t})}\psi^{*(\gamma )}(z)=e^{-\xi ({\bf t}_{\gamma}, z)}
\psi^{*(\gamma )}(z)e^{J({\bf t})},
\\ \\
e^{\bar J(\bar {\bf t})}\psi^{(\gamma )}(z)=e^{\xi 
(\bar {\bf t}_{\gamma}, z^{-1})}
\psi^{(\gamma )}(z)e^{\bar J({\bf t})}, \quad
e^{\bar J(\bar {\bf t})}\psi^{*(\gamma )}(z)=
e^{-\xi (\bar {\bf t}_{\gamma}, z^{-1})}
\psi^{*(\gamma )}(z)e^{\bar J(\bar {\bf t})},
\end{array}
\eeq
where
\beq\label{f5}
\xi ({\bf t}_{\gamma}, z)=\sum_{k\geq 1}t_{\gamma , k}z^k.
\eeq
Note also that 
\beq\label{f5a}
J({\bf t})\bigl |{\bf n}\bigr > =\bigl <{\bf n}\bigr |
\bar J(\bar {\bf t})=0,
\eeq
so $\bigl <{\bf n}\bigr |e^{\bar J(\bar {\bf t})}=
\bigl <{\bf n}\bigr |$, $e^{J({\bf t})}\bigl |{\bf n}\bigr >=
\bigl |{\bf n}\bigr >$.

\section{The bilinear identity}

\subsection{The bilinear identity in the operator form}

The Clifford group element of the fermionic algebra 
has the general form
\beq\label{int1a}
g=\exp \left ( \sum_{\alpha , \beta}
\sum_{j,k}\Bigl ( A_{jk}^{(\alpha \beta )}
\psi^{(\alpha )}_{j}\psi^{*(\beta )}_{k}+
B_{jk}^{(\alpha \beta )}
\psi^{(\alpha )}_{j}\psi^{(\beta )}_{k}+
C_{jk}^{(\alpha \beta )}
\psi^{*(\alpha )}_{j}\psi^{*(\beta )}_{k}\Bigr )
\right )
\eeq
with some infinite matrices $A_{jk}^{(\alpha \beta )}$, 
$B_{jk}^{(\alpha \beta )}$, $C_{jk}^{(\alpha \beta )}$, i.e.,
it is exponent of a quadratic form in the fermionic
operators $\psi_{j}^{(\alpha )}$, 
$\psi_{j}^{*(\alpha )}$. 
A characteristic property of the Clifford group elements is the following 
operator bilinear identity:
\beq\label{f2}
\sum_{\gamma =1}^N \sum_{j\in \z}\Bigl (
\psi_{j}^{(\gamma )}g \otimes
\psi_{j}^{*(\gamma )}g +
\psi_{j}^{*(\gamma )}g \otimes
\psi_{j}^{(\gamma )}g \Bigr )
=\sum_{\gamma =1}^N \sum_{j\in \z}\Bigl (
g\psi_{j}^{(\gamma )}\otimes g \psi_{j}^{*(\gamma )}+
g\psi_{j}^{*(\gamma )}\otimes g \psi_{j}^{(\gamma )}\Bigr ).
\eeq
The proof can be found in Appendix A to the paper \cite{SZ24}. 
In terms of the free fermionic fields
the operator bilinear identity acquires the form
\beq\label{f3}
\begin{array}{c}
\displaystyle{
\sum_{\gamma =1}^N \mbox{res}\left [ \frac{dz}{z} \Bigl (
\psi^{(\gamma )}(z)g\otimes \psi^{*(\gamma )}(z)g
+\psi^{*(\gamma )}(z)g\otimes \psi^{(\gamma )}(z)g\Bigr )\right ]}
\\ \\
\displaystyle{
=
\sum_{\gamma =1}^N \mbox{res} \left [\frac{dz}{z} \Bigl ( 
g\psi^{(\gamma )}(z)\otimes g\psi^{*(\gamma )}(z)
+g\psi^{*(\gamma )}(z)\otimes g\psi^{(\gamma )}(z)\Bigr )\right ]},
\end{array}
\eeq
where the symbol $\mbox{res}$ is defined as 
$\displaystyle{\mbox{res}\, \Bigl (\sum_{k\in \z}a_kz^k dz \Bigr )
=a_{-1}}$.
This bilinear identity will serve below as a starting point for 
derivation of the integral bilinear relation for the 
tau-function of the multi-component Pfaff-Toda hierarchy.

\subsection{The generating 
bilinear functional relation for the tau-func\-tion}

Let ${\bf n}=\{n_1, \ldots , n_N\}$, 
$\bar {\bf n}=\{\bar n_1, \ldots , \bar n_N\}$,
${\bf n}'=\{n_1', \ldots , n_N'\}$, 
$\bar {\bf n}'=\{\bar n_1', \ldots , \bar n_N'\}$
be four sets of integer variables. 
Sandwiching both sides of
(\ref{f3}) between the states
$$
\bigl <{\bf n}\bigr |
e^{J({\bf t})}\otimes \bigl <{\bf n}'\bigr |
e^{J({\bf t}')}\quad \mbox{and} \quad
e^{-\bar J(\bar {\bf t})}\bigl |-\bar {\bf n}\bigr > \otimes
e^{-\bar J(\bar {\bf t}')}\bigl |-\bar {\bf n}'\bigr >,
$$
we get:
\beq\label{f4}
\begin{array}{l}
\displaystyle{
\sum_{\gamma}\oint_{C_{\infty}}\frac{dz}{z}
e^{\xi ({\bf t}_{\gamma}-{\bf t}_{\gamma}', z)}
\bigl <{\bf n}\bigr |
\psi^{(\gamma )}(z)e^{J({\bf t})}ge^{-\bar J(\bar {\bf t})}
\bigl |-\bar {\bf n}\bigr >  
\bigl <{\bf n}'\bigr |
\psi^{*(\gamma )}(z)e^{J({\bf t}')}ge^{-\bar J(\bar {\bf t}')}
\bigl |-\bar {\bf n}'\bigr >}
\\ \\
\displaystyle{
+\, \sum_{\gamma}\oint_{C_{\infty}}\frac{dz}{z}
e^{-\xi ({\bf t}_{\gamma}-{\bf t}_{\gamma}', z)}
\bigl <{\bf n}\bigr |
\psi^{*(\gamma )}(z)e^{J({\bf t})}ge^{-\bar J(\bar {\bf t})}
\bigl |-\bar {\bf n}\bigr >  
\bigl <{\bf n}'\bigr |
\psi^{(\gamma )}(z)e^{J({\bf t}')}ge^{-\bar J(\bar {\bf t}')}
\bigl |-\bar {\bf n}'\bigr >}
\\ \\
\displaystyle{
=\, \sum_{\gamma}\oint_{C_{0}}\frac{dz}{z}
e^{\xi (\bar {\bf t}_{\gamma}-\bar {\bf t}_{\gamma}', z^{-1})}
\bigl <{\bf n}\bigr |
e^{J({\bf t})}ge^{-\bar J(\bar {\bf t})}\psi^{(\gamma )}(z)
\bigl |-\bar {\bf n}\bigr >  
\bigl <{\bf n}'\bigr |
e^{J({\bf t}')}ge^{-\bar J(\bar {\bf t}')}\psi^{*(\gamma )}(z)
\bigl |-\bar {\bf n}'\bigr >}
\\ \\
\displaystyle{
+\, \sum_{\gamma}\oint_{C_{0}}\frac{dz}{z}
e^{-\xi (\bar {\bf t}_{\gamma}-\bar {\bf t}_{\gamma}', z^{-1})}
\bigl <{\bf n}\bigr |
e^{J({\bf t})}ge^{-\bar J(\bar {\bf t})}\psi^{*(\gamma )}(z)
\bigl |-\bar {\bf n}\bigr >  
\bigl <{\bf n}'\bigr |
e^{J({\bf t}')}ge^{-\bar J(\bar {\bf t}')}\psi^{(\gamma )}(z)
\bigl |-\bar {\bf n}'\bigr >},
\end{array}
\eeq
where the commutation relations (\ref{comm}) have been used.
The integration contours $C_{\infty}, C_0$ are small circles
around $\infty$ and $0$ respectively. 

In order to further transform equation (\ref{f4}),
we employ the multicomponent bosonization rules \cite{KL93}
\beq\label{f5b}
\begin{array}{l}
\bigl <{\bf n}\bigr |\psi^{(\gamma )}(z)=\epsilon_{\gamma}({\bf n})
z^{n_{\gamma}-1}\bigl < {\bf n}-{\bf e}_{\gamma}\bigr |
e^{-J([z^{-1}]_{\gamma})},
\\ \\
\bigl <{\bf n}\bigr |\psi^{*(\gamma )}(z)=\epsilon_{\gamma}({\bf n})
z^{-n_{\gamma}}\bigl < {\bf n}+{\bf e}_{\gamma}\bigr |
e^{J([z^{-1}]_{\gamma})},
\\ \\
\psi^{(\gamma )}(z) \bigl |{\bf n}\bigr >=\epsilon_{\gamma}({\bf n})
z^{n_{\gamma}}e^{\bar J([z]_{\gamma})}
\bigl | {\bf n}+{\bf e}_{\gamma}\bigr >,
\\ \\
\psi^{*(\gamma )}(z) \bigl |{\bf n}\bigr >=\epsilon_{\gamma}({\bf n})
z^{-n_{\gamma}+1}e^{-\bar J([z]_{\gamma})}
\bigl | {\bf n}-{\bf e}_{\gamma}\bigr >.
\end{array}
\eeq
Here
${\bf e}_{\gamma}$ is
the $N$-component vector whose $\gamma$-th component is 1 and all others
are zero,
$[z]_{\gamma}$ is the set ${\bf t}$
such that ${\bf t}_{\mu}=\{0,0, \ldots \}$ if
$\mu \neq \gamma$, ${\bf t}_{\gamma}=
\{z, z^2/2, z^3/3, \ldots \}$ and 
\beq\label{f51a}
\epsilon_{\gamma}({\bf n})=(-1)^{n_{\gamma +1}+\ldots +n_N}
\eeq
is a sign factor. With the help of these formulas, equation (\ref{f4})
acquires the form
$$
\begin{array}{l}
\displaystyle{
\sum_{\gamma}
\epsilon_{\gamma}({\bf n})\epsilon_{\gamma}({\bf n}')
\oint_{C_{\infty}}dz z^{n_{\gamma}-n_{\gamma}'-2}
e^{\xi ({\bf t}_{\gamma}-{\bf t}_{\gamma}', z)}}
\\ \\
\phantom{aaaaaaaaa}\displaystyle{
\times \bigl <{\bf n}\! 
-\! {\bf e}_{\gamma}\bigr |
e^{J({\bf t}-[z^{-1}]_{\gamma})}ge^{-\bar J(\bar {\bf t})}
\bigl |-\bar {\bf n}\bigr >  
\bigl <{\bf n}'\! +
\! {\bf e}_{\gamma}\bigr |
e^{J({\bf t}'+[z^{-1}]_{\gamma})}ge^{-\bar J(\bar {\bf t}')}
\bigl |-\bar {\bf n}'\bigr >}
\\ \\
+\, \displaystyle{
\sum_{\gamma}
\epsilon_{\gamma}({\bf n})\epsilon_{\gamma}({\bf n}')
\oint_{C_{\infty}}dz z^{n'_{\gamma}-n_{\gamma}-2}
e^{-\xi ({\bf t}_{\gamma}-{\bf t}_{\gamma}', z)}}
\\ \\
\phantom{aaaaaaaaa}\displaystyle{
\times \bigl <{\bf n}\!
+\! {\bf e}_{\gamma}\bigr |
e^{J({\bf t}+[z^{-1}]_{\gamma})}ge^{-\bar J(\bar {\bf t})}
\bigl |-\bar {\bf n}\bigr >  
\bigl <{\bf n}'\! -
\! {\bf e}_{\gamma}\bigr |
e^{J({\bf t}'-[z^{-1}]_{\gamma})}ge^{-\bar J(\bar {\bf t}')}
\bigl |-\bar {\bf n}'\bigr >}
\end{array}
$$
\beq\label{f4a}
\begin{array}{l}
=\, 
\displaystyle{
\sum_{\gamma}
\epsilon_{\gamma}(\bar {\bf n})\epsilon_{\gamma}(\bar {\bf n}')
\oint_{C_{0}}dz z^{\bar n_{\gamma}'-\bar n_{\gamma}}
e^{\xi (\bar {\bf t}_{\gamma}-\bar {\bf t}_{\gamma}', z^{-1})}
}
\\ \\
\phantom{aaaaaaaaa} \displaystyle{
\times \bigl <{\bf n}\bigr |
e^{J({\bf t})}ge^{-\bar J(\bar {\bf t}-[z]_{\gamma})}
\bigl |-\bar {\bf n}\! +\! {\bf e}_{\gamma}\bigr >  
\bigl <{\bf n}'\bigr |
e^{J({\bf t}')}ge^{-\bar J(\bar {\bf t}'+[z]_{\gamma})}
\bigl |-\bar {\bf n}'\! -
\! {\bf e}_{\gamma}\bigr >
}
\\ \\
+\,
\displaystyle{
\sum_{\gamma}
\epsilon_{\gamma}(\bar {\bf n})\epsilon_{\gamma}(\bar {\bf n}')
\oint_{C_{0}}dz z^{\bar n_{\gamma}-\bar n_{\gamma}'}
e^{-\xi (\bar {\bf t}_{\gamma}-\bar {\bf t}_{\gamma}', z^{-1})}}
\\ \\
\phantom{aaaaaaaaa}\displaystyle{
\times \bigl <{\bf n}\bigr |
e^{J({\bf t})}ge^{-\bar J(\bar {\bf t}+[z]_{\gamma})}
\bigl |-\bar {\bf n}\! 
-\! {\bf e}_{\gamma}\bigr >  
\bigl <{\bf n}'\bigr |
e^{J({\bf t}')}ge^{-\bar J(\bar {\bf t}'-[z]_{\gamma})}
\bigl |-\bar {\bf n}'\! +
\! {\bf e}_{\gamma}\bigr >}.
\end{array}
\eeq

We define 
the tau-function of the multi-component  
Pfaff-Toda hierarchy to be 
the expectation value
\beq\label{f1}
\tau ({\bf n}, \bar {\bf n}, {\bf t},
\bar {\bf t})=
\lbr {\bf n}\right | e^{J({\bf t})} g 
e^{-\bar J(\bar {\bf t})}\left |-\bar {\bf n}\rbr ,
\eeq
where
$g$ is an element of the Clifford group of general form (\ref{int1a}). 
Note that $$\lbr {\bf n}\right | e^{J({\bf t})} g
e^{-\bar J(\bar {\bf t})}\left |-\bar {\bf n}\rbr =0$$ 
unless $\displaystyle{\sum_{\alpha}n_{\alpha}-
\sum_{\alpha}\bar n_{\alpha}}$ is an even
number, so $\displaystyle{|{\bf n}|=\sum_{\alpha}n_{\alpha}}$
and $\displaystyle{|\bar {\bf n}|=\sum_{\alpha}\bar n_{\alpha}}$
should be of the same parity (both even or both odd numbers).
In this sense the discrete variables ${\bf n}$, $\bar {\bf n}$ are
not fully independent.

Using the definition of the tau-function, we can rewrite (\ref{f4a})
as an integral bilinear equation for the tau-function. To bring it
to a more convenient form, we change the integration variable $z\to z^{-1}$
in the integrals in the right-hand side. Then the equation becomes
\beq\label{main}
\begin{array}{l}
\displaystyle{
\sum_{\gamma}\epsilon_{\gamma}({\bf n})\epsilon_{\gamma}({\bf n}')
\oint_{C_{\infty}}\frac{dz}{z^2} 
z^{n_{\gamma}-n_{\gamma}'}
e^{\xi ({\bf t}_{\gamma}-{\bf t}_{\gamma}', z)}}
\\ \\
\phantom{aaaaaaaaaaa}\displaystyle{\times \, 
\tau ({\bf n}\! -\! {\bf e}_{\gamma}, \bar {\bf n}, 
{\bf t}\! -\! [z^{-1}]_{\gamma},
\bar {\bf t})\tau ({\bf n}'\! +\! {\bf e}_{\gamma}, \bar {\bf n}', 
{\bf t}'\! +\! [z^{-1}]_{\gamma},
\bar {\bf t}')}
\\ \\
+\, 
\displaystyle{
\sum_{\gamma}\epsilon_{\gamma}({\bf n})\epsilon_{\gamma}({\bf n}')
\oint_{C_{\infty}}\frac{dz}{z^2} 
z^{n_{\gamma}'-n_{\gamma}}
e^{-\xi ({\bf t}_{\gamma}-{\bf t}_{\gamma}', z)}}
\\ \\
\phantom{aaaaaaaaaaa}\displaystyle{\times \, 
\tau ({\bf n}\! +\! {\bf e}_{\gamma}, \bar {\bf n}, 
{\bf t}\! +\! [z^{-1}]_{\gamma},
\bar {\bf t})\tau ({\bf n}'\! -\! {\bf e}_{\gamma}, \bar {\bf n}', 
{\bf t}'\! -\! [z^{-1}]_{\gamma},
\bar {\bf t}')}
\\ \\
=\, 
\displaystyle{
\sum_{\gamma}\epsilon_{\gamma}(\bar {\bf n})
\epsilon_{\gamma}(\bar {\bf n}')
\oint_{C_{\infty}}\frac{dz}{z^2} 
z^{\bar n_{\gamma}-\bar n_{\gamma}'}
e^{\xi (\bar {\bf t}_{\gamma}-\bar {\bf t}_{\gamma}', z)}}
\\ \\
\phantom{aaaaaaaaaaa}\displaystyle{\times \, 
\tau ({\bf n}, \bar {\bf n}\! -\! {\bf e}_{\gamma}, 
{\bf t},
\bar {\bf t}\! -\! [z^{-1}]_{\gamma})
\tau ({\bf n}', \bar {\bf n}'\! +\! {\bf e}_{\gamma}, 
{\bf t}',
\bar {\bf t}'\! +\! [z^{-1}]_{\gamma})}
\\ \\
+\, 
\displaystyle{
\sum_{\gamma}\epsilon_{\gamma}(\bar {\bf n})
\epsilon_{\gamma}(\bar {\bf n}')
\oint_{C_{\infty}}\frac{dz}{z^2} 
z^{\bar n_{\gamma}'-\bar n_{\gamma}}
e^{-\xi (\bar {\bf t}_{\gamma}-\bar {\bf t}_{\gamma}', z)}}
\\ \\
\phantom{aaaaaaaaaaa}\displaystyle{\times \, 
\tau ({\bf n}, \bar {\bf n}\! +\! {\bf e}_{\gamma}, 
{\bf t},
\bar {\bf t}\! +\! [z^{-1}]_{\gamma})
\tau ({\bf n}', \bar {\bf n}'\! -\! {\bf e}_{\gamma}, 
{\bf t}',
\bar {\bf t}'\! -\! [z^{-1}]_{\gamma})}.
\end{array}
\eeq
It is valid for all ${\bf t}, \bar {\bf t}$, ${\bf t}', \bar {\bf t}'$
and ${\bf n}, \bar {\bf n}$,  ${\bf n}', \bar {\bf n}'$
such that 
\beq\label{par1}
|{\bf n}|-|\bar {\bf n}|\in 2\ZZ +1, \qquad
|{\bf n}'|-|\bar {\bf n}'|\in 2\ZZ +1,
\eeq
i.e.,
the numbers $|{\bf n}|$ and $|\bar {\bf n}|$ as well as
$|{\bf n}'|$ and $|\bar {\bf n}|'$ have different 
parities\footnote{In fact if this is not the case equation 
(\ref{main}) is still valid becoming the trivial identity $0=0$}.
The integration contour $C_{\infty}$ 
around $\infty$ is a big circle of radius
$R\to \infty$ such that all singularities 
coming from the powers of $z$ and the exponential functions
$e^{\xi ({\bf t}_{\gamma}-{\bf t}_{\gamma}', \, z)}$, 
$e^{\xi (\bar {\bf t}_{\gamma}-\bar {\bf t}_{\gamma}', \, z)}$
are outside it and all singularities 
coming from the $\tau$-factors are inside it (the 
$\tau$-factors as functions of $z$ are regular in some
neighborhood of infinity.) 
Note that equation (\ref{main}) has the following obvious 
symmetries:
\beq\label{sym1}
({\bf n}, {\bf t}, \bar {\bf n}, \bar {\bf t}),
({\bf n}', {\bf t}', \bar {\bf n}', \bar {\bf t}')
\longleftrightarrow
({\bf n}', {\bf t}', \bar {\bf n}', \bar {\bf t}'),
({\bf n}, {\bf t}, \bar {\bf n}, \bar {\bf t}),
\eeq
and
\beq\label{sym2}
({\bf n}, {\bf t}, {\bf n}', {\bf t}'),
(\bar {\bf n},\bar {\bf t}, \bar {\bf n}', \bar {\bf t}')
\longleftrightarrow
(\bar {\bf n},\bar {\bf t}, \bar {\bf n}', \bar {\bf t}'),
({\bf n}, {\bf t}, {\bf n}', {\bf t}').
\eeq

At $N=1$ equation (\ref{main}) coincides with
the equation for the tau-function of the one-component Pfaff-Toda
hierarchy obtained by Takasaki 
in \cite{Takasaki09}. For any $N\geq 1$, after setting 
$\bar {\bf n}'=\bar {\bf n}$, $\bar {\bf t}'=\bar {\bf t}$
in (\ref{main}) its right-hand side vanishes identically
and the rest becomes the integral bilinear
equation for the tau-function of the multi-component 
DKP hierarchy from \cite{SZ24}.
To make comparison with those papers easier, one should 
pass from ${\bf n}$, $\bar {\bf n}$ 
to new discrete variables ${\bf s}=\{s_1, \ldots , s_N\}$,
${\bf r}=\{r_1, \ldots , r_N\}$ and consider the array of tau-functions
\beq\label{f1a}
\tau_{\alpha \beta}
({\bf s}, {\bf r}, {\bf t},
\bar {\bf t})=\lbr {\bf s}\! +\! {\bf r}\! +\! {\bf e}_{\alpha}
\! -\! {\bf e}_{\beta}\right | e^{J({\bf t})} g 
e^{-\bar J(\bar {\bf t})}\left |{\bf s}\! -\! {\bf r}\rbr 
=\tau ({\bf s}\! +\! {\bf r}\! +\! {\bf e}_{\alpha}\! -\! 
{\bf e}_{\beta}, {\bf r}\!-\! {\bf s}, {\bf t}, \bar {\bf t})
\eeq
which can be regarded as an $N\times N$ 
matrix-valued tau-function with 
matrix elements $\tau_{\alpha \beta}$. Note that the new 
discrete variables ${\bf s}, {\bf r}$ are fully independent.
In these new variables equation (\ref{main}) acquires the form
\beq\label{maina}
\begin{array}{l}
\displaystyle{
\sum_{\gamma}\epsilon_{\alpha \gamma}\epsilon_{\beta \gamma}
\epsilon_{\alpha \gamma}({\bf s})\epsilon_{\alpha \gamma}({\bf r})
\epsilon_{\beta \gamma}({\bf s}')\epsilon_{\beta \gamma}({\bf r}')
\oint_{C_{\infty}}dz z^{s_{\gamma}+r_{\gamma}-s_{\gamma}'-r_{\gamma}'
+\delta_{\alpha \gamma}+\delta_{\beta \gamma}-2}
e^{\xi ({\bf t}_{\gamma}-{\bf t}_{\gamma}', z)}}
\\ \\
\phantom{aaaaaaaaaaaaaa}
\times \tau_{\alpha \gamma}({\bf s}, {\bf r}, {\bf t}-[z^{-1}]_{\gamma},
\bar {\bf t})
\tau_{\gamma \beta}({\bf s}', {\bf r}', {\bf t}'+[z^{-1}]_{\gamma},
\bar {\bf t}')
\\ \\
+\, \displaystyle{
\sum_{\gamma}\epsilon_{\alpha \gamma}\epsilon_{\beta \gamma}
\epsilon_{\alpha \gamma}({\bf s})\epsilon_{\alpha \gamma}({\bf r})
\epsilon_{\beta \gamma}({\bf s}')\epsilon_{\beta \gamma}({\bf r}')
\oint_{C_{\infty}}dz z^{-s_{\gamma}-r_{\gamma}+s_{\gamma}'+r_{\gamma}'
-\delta_{\alpha \gamma}-\delta_{\beta \gamma}-2}
e^{-\xi ({\bf t}_{\gamma}-{\bf t}_{\gamma}', z)}}
\\ \\
\phantom{aaaaaaaaaaa}
\times \tau_{\beta \gamma}({\bf s}'-{\bf e}_{\beta}, 
{\bf r}'-{\bf e}_{\beta}, {\bf t}'-[z^{-1}]_{\gamma},
\bar {\bf t}')
\tau_{\gamma \alpha}({\bf s}+{\bf e}_{\alpha}, 
{\bf r}+{\bf e}_{\alpha}, {\bf t}+[z^{-1}]_{\gamma},
\bar {\bf t})
\\ \\
\displaystyle{
=\, \sum_{\gamma}
\epsilon_{\alpha \gamma}({\bf s})\epsilon_{\alpha \gamma}({\bf r})
\epsilon_{\beta \gamma}({\bf s}')\epsilon_{\beta \gamma}({\bf r}')
\oint_{C_{\infty}}dz z^{s_{\gamma}'-r_{\gamma}'-s_{\gamma}+r_{\gamma}-2}
e^{\xi (\bar {\bf t}_{\gamma}-\bar {\bf t}_{\gamma}', z)}}
\\ \\
\phantom{aaaaaaaaaaaaaa}
\times \tau_{\alpha \gamma}({\bf s}+{\bf e}_{\gamma}, {\bf r}, {\bf t},
\bar {\bf t}-[z^{-1}]_{\gamma})
\tau_{\gamma \beta}({\bf s}'-{\bf e}_{\gamma}, {\bf r}', {\bf t}',
\bar {\bf t}'+[z^{-1}]_{\gamma})
\\ \\
+\, \displaystyle{
\sum_{\gamma}
\epsilon_{\alpha \gamma}({\bf s})\epsilon_{\alpha \gamma}({\bf r})
\epsilon_{\beta \gamma}({\bf s}')\epsilon_{\beta \gamma}({\bf r}')
\oint_{C_{\infty}}dz z^{s_{\gamma}-r_{\gamma}-s_{\gamma}'+r_{\gamma}'-2}
e^{-\xi (\bar {\bf t}_{\gamma}-\bar {\bf t}_{\gamma}', z)}}
\\ \\
\phantom{aaaaaaaaaaaaaa}
\times \tau_{\gamma \beta}({\bf s}', 
{\bf r}'-{\bf e}_{\gamma}, {\bf t}',
\bar {\bf t}'-[z^{-1}]_{\gamma})
\tau_{\alpha \gamma}({\bf s}, 
{\bf r}+{\bf e}_{\gamma}, {\bf t},
\bar {\bf t}+[z^{-1}]_{\gamma}),
\end{array}
\eeq
where the sign factor $\epsilon_{\alpha \gamma}({\bf s})$ is defined as
\beq\label{f51}
\epsilon_{\alpha \gamma}({\bf s})=\left \{
\begin{array}{l}
(-1)^{s_{\alpha +1}+ \ldots + s_{\gamma}} \quad 
\phantom{a}\mbox{if $\alpha < \gamma$},
\\ 
1 \phantom{aaaaaaaaaaaaaa}\mbox{if $\alpha = \gamma$},
\\ 
-(-1)^{s_{\gamma +1}+ \ldots + s_{\alpha}} \quad 
\mbox{if $\alpha > \gamma$}
\end{array}
\right.
\eeq
and
the sign factor $\epsilon_{\alpha \beta}=\epsilon_{\alpha \beta}({\bf 0})$ is
$\epsilon_{\alpha \beta}=1$ if $\alpha \leq \beta$, 
$\epsilon_{\alpha \beta}=-1$ if $\alpha > \beta$.
After setting $\bar {\bf t}=\bar {\bf t}'$, ${\bf s}={\bf r}
+{\bf n}_0$,
${\bf s}'={\bf r}' +{\bf n}_0$
the right hand side vanishes and the rest becomes the integral bilinear
equation for the tau-function 
$$
\tau_{\alpha \beta}^{\rm DKP}({\bf r}, {\bf t})=\tau_{\alpha \beta}
({\bf r}+{\bf n}_0, {\bf r}, {\bf t}, \bar {\bf t})=
\bigl <2{\bf r}+{\bf e}_{\alpha}-{\bf e}_{\beta}+{\bf n}_0\bigr |
e^{J({\bf t})}g e^{-J(\bar {\bf t})}\bigl |{\bf n}_0\bigr >
$$
of the multi-component DKP hierarchy
obtained in the paper \cite{SZ24}.
The independent variables
are ${\bf r}, {\bf t}$ while $\bar {\bf t}$
and ${\bf n}_0$
do not participate in the equation and the tau-function depends 
on them as on parameters\footnote{In \cite{SZ24} 
${\bf n}_0$ and $\bar {\bf t}$ were the sets of zeros.}. 
Similarly, set ${\bf t}={\bf t}'$,
${\bf s}=\bar {\bf s}-{\bf e}_{\alpha}$,
${\bf s}'=\bar {\bf s}'+{\bf e}_{\beta}$, $\bar {\bf s}={\bf n}_0
-\bar {\bf r}$,
$\bar {\bf s}'={\bf n}_0-\bar {\bf r}'$ in (\ref{main}), 
then the left hand side
vanishes while the rest becomes the integral bilinear equation
for the tau-function
$$
\bar \tau_{\alpha \beta}^{\rm DKP}(\bar {\bf r}, \bar {\bf t})=
\tau_{\alpha \beta}
(-\bar {\bf r}-{\bf e}_{\alpha}+{\bf e}_{\beta}+{\bf n}_0, \bar {\bf r}, 
{\bf t}, \bar {\bf t})=
\bigl < {\bf n}_0\bigr |
e^{J({\bf t})}g e^{-\bar J(\bar {\bf t})}\bigl |
{\bf n}_0+{\bf e}_{\beta}-{\bf e}_{\alpha}-2\bar {\bf s}\bigr >
$$
as a function of the variables $\bar {\bf r}, \bar {\bf t}$.

In what follows we will work with the variables ${\bf n}$, $\bar {\bf n}$
and deal with equation (\ref{main}) rather than (\ref{maina})
since the structure of the former is much more clear and suggestive.
The only advantage of the latter equation 
is that the discrete variables ${\bf s}$, ${\bf r}$
are independent but this is achieved for a too big price of making
the symmetries (\ref{sym1}), (\ref{sym2}) implicit.

For the calculations below it is useful to put
$$
{\bf n}'={\bf n}+{\bf k}, \qquad \bar {\bf n}'=\bar {\bf n}+\bar {\bf k}.
$$
Note that the parity restrictions imply that $|{\bf k}|$ and
$|\bar {\bf k}|$ must be of the same parity, i.e.,
$$
|{\bf k}|-|\bar {\bf k}|\in 2 \ZZ .
$$
In this notation,
equation (\ref{main}) acquires the form
$$
\begin{array}{l}
\displaystyle{
\sum_{\gamma}\epsilon_{\gamma}({\bf k})
\oint_{C_{\infty}}dz
z^{-k_{\gamma}-2}
e^{\xi ({\bf t}_{\gamma}-{\bf t}_{\gamma}', z)}}
\\ \\
\phantom{aaaaaaaaaaa}\displaystyle{\times \, 
\tau ({\bf n}\! -\! {\bf e}_{\gamma}, \bar {\bf n}, 
{\bf t}\! -\! [z^{-1}]_{\gamma},
\bar {\bf t})\tau ({\bf n}\! +\!{\bf k}\! +\! 
{\bf e}_{\gamma}, \bar {\bf n}\! +\! \bar {\bf k}, 
{\bf t}'\! +\! [z^{-1}]_{\gamma},
\bar {\bf t}')}
\\ \\
+\, 
\displaystyle{
\sum_{\gamma}\epsilon_{\gamma}({\bf k})
\oint_{C_{\infty}}dz
z^{k_{\gamma}-2}
e^{-\xi ({\bf t}_{\gamma}-{\bf t}_{\gamma}', z)}}
\\ \\
\phantom{aaaaaaaaaaa}\displaystyle{\times \, 
\tau ({\bf n}\! +\! {\bf e}_{\gamma}, \bar {\bf n}, 
{\bf t}\! +\! [z^{-1}]_{\gamma},
\bar {\bf t})\tau ({\bf n}\! +\! {\bf k} 
\! -\! {\bf e}_{\gamma}, \bar {\bf n}+\bar {\bf k}, 
{\bf t}'\! -\! [z^{-1}]_{\gamma},
\bar {\bf t}')}
\end{array}
$$
\beq\label{main1}
\begin{array}{l}
=\, 
\displaystyle{
\sum_{\gamma}\epsilon_{\gamma}(\bar {\bf k})
\oint_{C_{\infty}}dz 
z^{-\bar k_{\gamma}-2}
e^{\xi (\bar {\bf t}_{\gamma}-\bar {\bf t}_{\gamma}', z)}}
\\ \\
\phantom{aaaaaaaaaaa}\displaystyle{\times \, 
\tau ({\bf n}, \bar {\bf n}\! -\! {\bf e}_{\gamma}, 
{\bf t},
\bar {\bf t}\! -\! [z^{-1}]_{\gamma})
\tau ({\bf n}+{\bf k}, \bar {\bf n}\! +\! \bar {\bf k}
\! +\! {\bf e}_{\gamma}, 
{\bf t}',
\bar {\bf t}'\! +\! [z^{-1}]_{\gamma})}
\\ \\
+\, 
\displaystyle{
\sum_{\gamma}\epsilon_{\gamma}(\bar {\bf k})
\oint_{C_{\infty}}dz
z^{\bar k_{\gamma}-2}
e^{-\xi (\bar {\bf t}_{\gamma}-\bar {\bf t}_{\gamma}', z)}}
\\ \\
\phantom{aaaaaaaaaaa}\displaystyle{\times \, 
\tau ({\bf n}, \bar {\bf n}\! +\! {\bf e}_{\gamma}, 
{\bf t},
\bar {\bf t}\! +\! [z^{-1}]_{\gamma})
\tau ({\bf n}+{\bf k}, \bar {\bf n}\! +\! \bar {\bf k}
\! -\! {\bf e}_{\gamma}, 
{\bf t}',
\bar {\bf t}'\! -\! [z^{-1}]_{\gamma})}.
\end{array}
\eeq

To clarify the structure of equation (\ref{main}) for the further
analysis, it is instructive to rewrite it in the short-hand notation
introduced below. Let $\hat {\bf t}_{\alpha}$, 
$\hat {\bar {\bf t}}_{\alpha}$ be the sets of
times (\ref{times}) extended by the ``zeroth times'' $t_{\alpha ,0}=n_{\alpha}$,
$\bar t_{\alpha ,0}=\bar n_{\alpha}$:
\beq\label{times1}
\hat {\bf t}_{\alpha}=\{ n_{\alpha}, t_{\alpha ,1}, t_{\alpha ,2}, \ldots \},
\qquad
\hat {\bar {\bf t}}_{\alpha}=\{\bar n_{\alpha}, 
\bar t_{\alpha ,1}, \bar t_{\alpha ,2}, \ldots \}, \quad \alpha =1, \ldots , N,
\eeq
the full extended sets of times are then
$$
\hat {\bf t}=\{\hat {\bf t}_{1}, \hat {\bf t}_{2}, 
\ldots , \hat {\bf t}_{N}\},
\qquad
\hat {\bar {\bf t}}=\{\hat {\bar {\bf t}}_{1}, \hat {\bar {\bf t}}_{2}, 
\ldots , \hat {\bar {\bf t}}_{N}\}.
$$
For these extended sets of times,
we introduce the generalized symbol $[z^{-1}]_{\gamma}^{\circ}$:
\beq\label{par2}
\hat {\bf t}_{\gamma}\pm [z^{-1}]^{\circ}_{\gamma}=\{n_{\gamma}\pm 1,
t_{\alpha ,1}\pm z^{-1}, t_{\alpha ,2}\pm z^{-1}/2, 
t_{\alpha ,3}\pm z^{-1}/3, \ldots \}
\eeq
and put
\beq\label{par3}
\xi (\hat {\bf t}_{\gamma}, z)=n_{\gamma}\log z+\sum_{k\geq 1}
t_{\gamma ,k}z^{k}.
\eeq
The tau-function is
\beq\label{times3}
\tau (\hat {\bf t}, \hat {\bar {\bf t}})=\tau ({\bf n}, \bar {\bf n},
{\bf t}, \bar {\bf t}).
\eeq
In this notation, equation (\ref{main}) acquires the following more compact
form:
\beq\label{main2}
\begin{array}{l}
\displaystyle{
\sum_{\gamma =1}^N \epsilon_{\gamma}({\bf n}\! -\! {\bf n}')
\oint_{C_{\infty}}\frac{dz}{z^2} \left [
e^{\xi (\hat {\bf t}_{\gamma}-\hat {\bf t}'_{\gamma}, z)}
\tau (\hat {\bf t}-[z^{-1}]^{\circ}_{\gamma}, \hat {\bar {\bf t}})
\tau (\hat {\bf t}'+[z^{-1}]^{\circ}_{\gamma}, \hat {\bar {\bf t}'})\right. }
\\ \\
\phantom{aaaaaaaaaaaaaaaaaaaaaaa}\displaystyle{\left. 
\vphantom{\oint_{C_{\infty}}\frac{dz}{z^2} }
+e^{-\xi (\hat {\bf t}_{\gamma}-\hat {\bf t}'_{\gamma}, z)}
\tau (\hat {\bf t}+[z^{-1}]^{\circ}_{\gamma}, \hat {\bar {\bf t}})
\tau (\hat {\bf t}'-[z^{-1}]^{\circ}_{\gamma}, \hat {\bar {\bf t}'})
\right ]}
\\ \\
=\, \displaystyle{
\sum_{\gamma =1}^N \epsilon_{\gamma}(\bar {\bf n}\! -\! \bar {\bf n}')
\oint_{C_{\infty}}\frac{dz}{z^2} \left [
e^{\xi (\hat {\bar {\bf t}}_{\gamma}-\hat {\bar {\bf t}'}_{\gamma}, z)}
\tau (\hat {\bf t}, \hat {\bar {\bf t}}-[z^{-1}]^{\circ}_{\gamma})
\tau (\hat {\bf t}', \hat {\bar {\bf t}'}+[z^{-1}]^{\circ}_{\gamma})\right. }
\\ \\
\phantom{aaaaaaaaaaaaaaaaaaaaaaa}\displaystyle{\left. 
\vphantom{\oint_{C_{\infty}}\frac{dz}{z^2} }
e^{-\xi (\hat {\bar {\bf t}}_{\gamma}-\hat {\bar {\bf t}'}_{\gamma}, z)}
\tau (\hat {\bf t}, \hat {\bar {\bf t}}+[z^{-1}]^{\circ}_{\gamma})
\tau (\hat {\bf t}', \hat {\bar {\bf t}'}-[z^{-1}]^{\circ}_{\gamma})\right ]}.
\end{array}
\eeq
This form helps to clarify the structure but it turns out that
for practical calculations the more detailed form (\ref{main}) or
(\ref{main1}) is more convenient.

\section{Hirota-Miwa functional equations}

Our next goal is to derive bilinear functional relations for the
tau-function which are obtained as corollaries of the general equation
(\ref{main}) after choosing the variables in some special ways
such that the integrals can be explicitly 
evaluated by means of residue calculus. The functional relations
obtained in this way are known as equations of the Hirota-Miwa type
(sometimes they are called differential or difference Fay
identities\footnote{We do not think that such a name is appropriate
because these equations become what is called Fay identities 
only for the class of algebraic-geometrical solutions in terms of
Riemann theta-functions.}).

From (\ref{main1}) many various equations of 
the Hirota-Miwa type follow. 
Our goal is to write down 
all equations of the Hirota-Miwa type such that
\begin{itemize}
\item[A)] They contain not more than four different 
bilinear terms, 
\item[B)] They do not contain any derivatives 
(with respect to the times from the sets 
${\bf t}_1$, ${\bf t}_2$, etc.). 
\end{itemize}
Hereafter, we refer to these requirements 
as conditions A and B respectively.

\subsection{The simplest examples}

To proceed, we should first choose the values ${\bf t}-{\bf t}'$
and $\bar {\bf t}-\bar {\bf t}'$ in such a way that the integrals
could be evaluated by means of residue calculus. 
There are many different possibilities to do that. 
Before addressing the general form of them, let us consider 
one of the simplest examples corresponding to the case when
the multi-component hierarchy is reduced to the $1$-component one. 

We begin with the choice
$$
({\bf k}, \bar {\bf k})=(-{\bf e}_{\alpha}, {\bf e}_{\alpha}),
\quad 
\left \{\begin{array}{l}
{\bf t}-{\bf t}' =[a^{-1}]_{\alpha}+[b^{-1}]_{\alpha}, 
\\ 
\bar {\bf t}-\bar {\bf t}'=0
\end{array} 
\right. 
$$
in (\ref{main1}), where $a, b \in \CC$ are arbitrary parameters
belonging to a neighborhood of infinity.
It is a particular case of the Miwa change of variables.
In this case only the variables corresponding to the $\alpha$th
component participate in the equations, and the $N$-component
Pfaff-Toda lattice reduces to the 1-component hierarchy
(for each $\alpha =1, \ldots , N$)
introduced in \cite{Takasaki09}. 

For this simple example, 
let us explain in some details 
how to derive equations of the Hirota-Miwa type
from (\ref{main1}). Most of the details are common for 
more complicated cases given later. 
We have:
$$
e^{\xi ({\bf t}_{\gamma} -{\bf t}'_{\gamma}, z)}=
\frac{a\, b}{(a-z)(b-z)} \quad \mbox{for $\gamma =\alpha$} \;\;
\mbox{and $1$ otherwise},
$$
so equation (\ref{main1}) converts into
\beq\label{sub1-4}
\begin{array}{c}
\displaystyle{\hspace{-1cm}
\oint_{C_{\infty}} \! dz  \frac{ab}{z(a-z)(b-z)}\, 
\tau \Bigl ({\bf n}-{\bf e}_{\alpha}, \bar {\bf n},
{\bf t}-[z^{-1}]_{\alpha}, \bar {\bf t}\Bigr )
\tau \Bigl ({\bf n}, \bar {\bf n}+{\bf e}_{\alpha},
{\bf t}'+[z^{-1}]_{\alpha}, \bar {\bf t}\Bigr )}
\\ \\
\displaystyle{+
\oint_{C_{\infty}} \! dz  \frac{(a-z)(b-z)}{z^3 \, ab}\, 
\tau \Bigl ({\bf n}+{\bf e}_{\alpha}, \bar {\bf n},
{\bf t}+[z^{-1}]_{\alpha}, \bar {\bf t}\Bigr )
\tau \Bigl ({\bf n}-2{\bf e}_{\alpha}, 
\bar {\bf n}+{\bf e}_{\alpha},
{\bf t}'-[z^{-1}]_{\alpha}, \bar {\bf t}\Bigr )}
\\ \\
\displaystyle{\, =
\oint_{C_{\infty}}\!\!  dz \, z^{-1} 
\tau \Bigl ({\bf n}, \bar {\bf n}+{\bf e}_{\alpha},
{\bf t}, \bar {\bf t}+[z^{-1}]_{\alpha}\Bigr )
\tau \Bigl ({\bf n}-{\bf e}_{\alpha}, 
\bar {\bf n},
{\bf t}', \bar {\bf t}-[z^{-1}]_{\alpha}\Bigr )},
\end{array}
\eeq
where only nonzero terms of the sums in (\ref{main1}) are shown.
The integrals can be calculated by residue calculus.
The expression under the first integral has simple poles at
$z=a,b$. Calculating the residues, one should take into account
that the both points $a,b$ are {\it outside} the contour, so 
the residues at these points should be taken with the sign
minus; and there might be also a contribution from $\infty$,
which is zero for this case. To the second and third integrals
only the residues at $\infty$ contribute. After some simple 
transformations and a shift of the variables, one obtains
the following equation:
\beq\label{eq1-1}
\begin{array}{l}
a\tau \Bigl ({\bf n}, \bar {\bf n}, {\bf t}+[a^{-1}]_{\alpha},
\bar {\bf t}\Bigr )
\tau \Bigl ({\bf n}+{\bf e}_{\alpha}, 
\bar {\bf n}+{\bf e}_{\alpha}, {\bf t}+[b^{-1}]_{\alpha},
\bar {\bf t}\Bigr )
\\ \\ \phantom{aaa}
-\, b\tau \Bigl ({\bf n}, \bar {\bf n}, {\bf t}+[b^{-1}]_{\alpha},
\bar {\bf t}\Bigr )
\tau \Bigl ({\bf n}+{\bf e}_{\alpha}, 
\bar {\bf n}+{\bf e}_{\alpha}, {\bf t}+[a^{-1}]_{\alpha},
\bar {\bf t}\Bigr )
\\ \\ \phantom{aaaaaaaa}
=\, (a-b)
\tau \Bigl ({\bf n}+{\bf e}_{\alpha}, 
\bar {\bf n}+{\bf e}_{\alpha}, {\bf t}+[a^{-1}]_{\alpha}
+[b^{-1}]_{\alpha}, \bar {\bf t}\Bigr )
\tau \Bigl ({\bf n}, \bar {\bf n}, {\bf t},\bar {\bf t}\Bigr )
\\ \\ \phantom{aaaaaaaaaaa}
+\, (a^{-1}-b^{-1})
\tau \Bigl ({\bf n}+2{\bf e}_{\alpha}, \bar {\bf n}, {\bf t}+[a^{-1}]_{\alpha}+[b^{-1}]_{\alpha}, \bar {\bf t}\Bigr )
\tau \Bigl ({\bf n}-{\bf e}_{\alpha}, 
\bar {\bf n}+{\bf e}_{\alpha}, {\bf t}, \bar {\bf t}\Bigr ).
\end{array}
\eeq
One can proceed in a similar way with another choice of the vectors
${\bf k}, \bar {\bf k}$:
$({\bf k}, \bar {\bf k})=(-{\bf e}_{\alpha}, -{\bf e}_{\alpha})$
and obtain the equation
\beq\label{eq1-2}
\begin{array}{l}
a\tau \Bigl ({\bf n}, \bar {\bf n}+{\bf e}_{\alpha}, 
{\bf t}+[a^{-1}]_{\alpha}, \bar {\bf t}\Bigr )
\tau \Bigl ({\bf n}+{\bf e}_{\alpha}, 
\bar {\bf n}, {\bf t}+[b^{-1}]_{\alpha}, \bar {\bf t}\Bigr )
\\ \\ \phantom{aaa}
-\, b\tau \Bigl ({\bf n}, \bar {\bf n}+{\bf e}_{\alpha}, 
{\bf t}+[b^{-1}]_{\alpha}, \bar {\bf t}\Bigr )
\tau \Bigl ({\bf n}+{\bf e}_{\alpha}, 
\bar {\bf n}, {\bf t}+[a^{-1}]_{\alpha},
\bar {\bf t}\Bigr )
\\ \\ \phantom{aaaaaaaa}
=\, (a-b)
\tau \Bigl ({\bf n}+{\bf e}_{\alpha}, 
\bar {\bf n}, {\bf t}+[a^{-1}]_{\alpha}+[b^{-1}]_{\alpha}, 
\bar {\bf t}\Bigr )
\tau \Bigl ({\bf n}, \bar {\bf n}+{\bf e}_{\alpha}, 
{\bf t}, \bar {\bf t}\Bigr )
\\ \\ \phantom{aaaaaaaaaaa}
+\, (a^{-1}-b^{-1})
\tau \Bigl ({\bf n}+2{\bf e}_{\alpha}, \bar {\bf n}
+{\bf e}_{\alpha}, 
{\bf t}+[a^{-1}]_{\alpha}+[b^{-1}]_{\alpha}, \bar {\bf t}\Bigr )
\tau \Bigl ({\bf n}-{\bf e}_{\alpha}, 
\bar {\bf n}, {\bf t}, \bar {\bf t}\Bigr ).
\end{array}
\eeq
After a proper linear change
of variables, these equations coincide with those obtained by 
Takasaki in \cite{Takasaki09} for the 1-component Pfaff-Toda hierarchy.

\subsection{The general case}

To approach the general case, one should employ the Miwa change of
variables (first introduced by Miwa in \cite{Miwa82}) 
in its general form, i.e., put 
$$
{\bf t}-{\bf t}'=\sum_i [a_i^{-1}]_{\alpha_i}-
\sum_i [b_i^{-1}]_{\beta_i}
$$
for the times ${\bf t}$ and similarly for the times $\bar {\bf t}$.
The sums here are assumed to be finite and
$\alpha_i, \beta_i$ are arbitrary indices from the set
$\{1, \ldots , N\}$. 
After this substitution, the integrals in (\ref{main1})
can be calculated by taking residues at poles at the points $a_i, b_i$
(the poles are simple if all of them are distinct). Besides, depending
on the choice of ${\bf n}-{\bf n}'$, there may be
a nonzero residue at infinity. Note, however, that for the particular
choice of ${\bf n}-{\bf n}'$ corresponding to the extension of the Miwa
change to the extended set of times (\ref{times1}) of the form
$$
\hat {\bf t}-\hat {\bf t}'=\sum_i [a_i^{-1}]^{\circ}_{\alpha_i}-
\sum_i [b_i^{-1}]^{\circ}_{\beta_i},
$$
the residue at infinity is always zero. In slightly different 
words, this means
that each shift ${\bf t}_{\alpha}\to {\bf t}_{\alpha}\pm [a^{-1}]_{\alpha}$
should be supplemented by the corresponding shift $n_{\alpha}\to n_{\alpha}
\pm 1$ (and similarly for the bar-times).

Keeping this in mind, we are going to consider 
the substitutions of the following general form:
\beq\label{m1}
\begin{array}{ll}
\displaystyle{
{\bf n}-{\bf n}' =-{\bf k}=\sum_{i=1}^{L^+}{\bf e}_{\alpha_i}-
\sum_{k=1}^{L^-}{\bf e}_{\beta_k},  \quad }
& \displaystyle{
\bar {\bf n}-\bar {\bf n}' = -\bar {\bf k}=\sum_{i=1}^{R^+}{\bf e}_{\bar \alpha_i}-
\sum_{k=1}^{R^-}{\bf e}_{\bar \beta_k}, \quad }
\\ & \\
\displaystyle{
{\bf t}-{\bf t}' =-{\bf T}=\sum_{i=1}^{L^+}[z_i^{-1}]_{\alpha_i}-
\sum_{k=1}^{L^-}[\zeta_k^{-1}]_{\beta_k}, \quad }
& \displaystyle{
\bar {\bf t}-\bar {\bf t}' =-\bar {\bf T}=
\sum_{i=1}^{R^+}[\tilde z_i^{-1}]_{\bar \alpha_i}-
\sum_{k=1}^{R^-}[\tilde \zeta_k^{-1}]_{\bar \beta_k}, \quad }.
\end{array}
\eeq
Here $\alpha_i, \beta_k, \bar \alpha_i, \bar \beta_k$ are arbitrary indices
from the set $\{1, \ldots , N\}$ (they may enter with multiplicities, i.e.,
the cases when $\alpha_i =\alpha_j$ for $i\neq j$ are allowed), and
$z_i , \zeta_k , \tilde z_i, \tilde \zeta_k \in \CC$ are arbitrary 
parameters belonging to a neighborhood of infinity (and, again, we allow the
cases when some of them coincide). If some of these points tend to infinity,
a nonzero residue at infinity arises. As is easy to see, the two procedures
(taking residue at some $z_i$ and tending $z_i$ to infinity,  
obtaining a non-zero residue there) commute.
Therefore, without any loss of generality, we can start from the case
when all complex variables are finite and, if necessary, tend 
some of them to infinity afterwards.
It is important to note that the parity restriction leads to the
following condition on possible values of $L^{\pm}, R^{\pm}$:
\beq\label{m2}
|L^+-L^-|-|R^+-R^-|\in 2\ZZ .
\eeq
We call the case when $z_i , \zeta_k , \tilde z_i, \tilde \zeta_k$  
and $\alpha_i, \beta_k, \bar \alpha_i, \bar \beta_k$
are all distinct {\it non-degenerate}. In this case all poles are simple
and there are no contributions from infinity. 

The key formula necessary for reducing the integrals in 
(\ref{main}) to finite sums of residues is simply 
\beq\label{m3}
z\, e^{\xi ([z_i], z)}=z\, e^{-\log (1-z/z_i)}=\frac{zz_i}{z_i-z}=
\frac{1}{z^{-1}-z_i^{-1}}.
\eeq
For the general substitution the corresponding factor is
$$
z^{n_{\gamma}-n'_{\gamma}}\, e^{\xi ({\bf t}_{\gamma}-{\bf t}'_{\gamma}, z)}
=\prod_{i=1}^{L^+}(z^{-1}-z_i^{-1})^{-\delta_{\alpha_i \gamma}}
\prod_{k=1}^{L^-}(z^{-1}-\zeta_k^{-1})^{\delta_{\beta_k \gamma}}.
$$
A similar factorization holds also for the sign factors:
\beq\label{m4}
\epsilon_{\gamma}({\bf n}- {\bf n}')=\prod_{i=1}^{L^+}
\epsilon_{\alpha_i \gamma}
\prod_{k=1}^{L^-}
\epsilon_{\beta_k \gamma}.
\eeq
It is convenient to introduce the function
\beq\label{m5}
E_{\alpha \beta}(z_i, z_j)=\epsilon_{\alpha \beta}
(z_i^{-1}-z_j^{-1})^{\delta_{\alpha \beta}}.
\eeq
Obviously, $E_{\beta \alpha}(z_j, z_i)=-E_{\alpha \beta}(z_i, z_j)$.

A direct calculation of residues in (\ref{main}) after the substitution 
(\ref{m1}) leads to the following general Hirota-Miwa equation:
\beq\label{genHM}
\begin{array}{l}
\displaystyle{
\sum_{s=1}^{L^+}
\prod_{{\scriptsize \begin{array}{l}i=1\\ i\neq s \end{array}}}^{L^+}
E^{-1}_{\alpha_i \alpha_s}(z_s, z_i)
\prod_{k=1}^{L^-}
E_{\beta_k \alpha_s}(z_s, \zeta_k)}
\\ \\
\phantom{aaaaaaa}\displaystyle{
\times \, 
\tau \Bigl ({\bf n}-{\bf e}_{\alpha_s}, \bar {\bf n}, {\bf t}-[z_s^{-1}]_{\alpha_s},
\bar {\bf t}\Bigr )
\tau \Bigl ({\bf n}+{\bf k}+{\bf e}_{\alpha_s}, \bar {\bf n}+\bar {\bf k}, 
{\bf t}+{\bf T}+[z_s^{-1}]_{\alpha_s},
\bar {\bf t}+\bar {\bf T}\Bigr )}
\\ \\
\displaystyle{
+\sum_{s=1}^{L^-}
\prod_{{\scriptsize \begin{array}{l}k=1\\ k\neq s \end{array}}}^{L^-}
E^{-1}_{\beta_k \beta_s}(\zeta_s, \zeta_k)
\prod_{i=1}^{L^+}
E_{\alpha_i \beta_s}(\zeta_s, z_i)}
\\ \\
\phantom{aaaaaaa}\displaystyle{
\times \, 
\tau \Bigl ({\bf n}+{\bf e}_{\beta_s}, \bar {\bf n}, 
{\bf t}+[\zeta_s^{-1}]_{\beta_s},
\bar {\bf t}\Bigr )
\tau \Bigl ({\bf n}+{\bf k}-{\bf e}_{\beta_s}, \bar {\bf n}+\bar {\bf k}, 
{\bf t}+{\bf T}-[\zeta_s^{-1}]_{\beta_s},
\bar {\bf t}+\bar {\bf T}\Bigr )}
\end{array}
\eeq
$$
\begin{array}{l}
\displaystyle{
=\sum_{s=1}^{R^+}
\prod_{{\scriptsize \begin{array}{l}j=1\\ j\neq s \end{array}}}^{R^+}
E^{-1}_{\bar \alpha_j \bar \alpha_s}(\tilde z_s, \tilde z_j)
\prod_{l=1}^{R^-}
E_{\bar \beta_l \bar \alpha_s}(\tilde z_s, \tilde \zeta_l)}
\\ \\
\phantom{aaaaaaa}\displaystyle{
\times \, 
\tau \Bigl ({\bf n}, \bar {\bf n}-{\bf e}_{\bar \alpha_s}, 
{\bf t},
\bar {\bf t}-[\tilde z_s^{-1}]_{\bar \alpha_s}\Bigr )
\tau \Bigl ({\bf n}+{\bf k}, \bar {\bf n}+\bar {\bf k}+{\bf e}_{\bar \alpha_s}, 
{\bf t}+{\bf T},
\bar {\bf t}+\bar {\bf T}+[\tilde z_s^{-1}]_{\bar \alpha_s}\Bigr )}
\\ \\
\displaystyle{
+\sum_{s=1}^{R^-}
\prod_{{\scriptsize \begin{array}{l}l=1\\ l\neq s \end{array}}}^{R^-}
E^{-1}_{\bar \beta_k \bar \beta_s}(\tilde \zeta_s, \tilde \zeta_k)
\prod_{j=1}^{R^+}
E_{\bar \alpha_j \bar \beta_s}(\tilde \zeta_s, \tilde z_j)}
\\ \\
\phantom{aaaaaaa}\displaystyle{
\times \, 
\tau \Bigl ({\bf n}, \bar {\bf n}+{\bf e}_{\bar \beta_s}, 
{\bf t},
\bar {\bf t}+[\tilde \zeta_s^{-1}]_{\bar \beta_s}\Bigr )
\tau \Bigl ({\bf n}+{\bf k}, \bar {\bf n}+\bar {\bf k}-{\bf e}_{\bar \beta_s}, 
{\bf t}+{\bf T},
\bar {\bf t}+\bar {\bf T}-[\tilde \zeta_s^{-1}]_{\bar \beta_s}\Bigr )}.
\end{array}
$$
It is the most general non-degenerate Hirota-Miwa equation
(functional relation) for the tau-function. 
It contains 
$$
M=L^+ +L^- + R^+ +R^- 
$$
bilinear terms, each of which is product of two tau-functions with
various shifts of the arguments. The coefficients are rational functions
of $z_i , \zeta_k , \tilde z_i, \tilde \zeta_k$. We will call it
the (non-degenerate) $M$-point relation, according to the total
number of the points. So, the number of terms in non-degenerate
relations coincides with the number of points. Besides, in the non-degenerate 
case, when all the points are distinct, the relation does not contain any
derivatives of the tau-function with respect to the continuous times.
Note, however, that this general $M$-point relation still holds in 
degenerate cases, when some of the points merge or tend to infinity.
In such cases, some terms of the general equation (\ref{genHM}) may
become singular, if one considers them separately. In the full
expression, the singularities 
can be resolved, and, as a result, 
derivatives of the tau-functions with respect to 
continuous times arise 
in this way.

Note also that if some points in the non-degenerate $M$-point relation 
tend to infinity, the number
of points (i.e., the number of the remaining free parameters), $m$, 
in such degenerate relations becomes 
strictly less than $M$. We will call them $m$-point relations (not
necessarily non-degenerate), and consider them as reductions of some
non-degenerate relations for some $M> m$. (The number of
different bilinear terms in them is still $M$.) 
Below we present some details
of the simplest and most important particular case $M=4$, $m=2$.

\subsection{Non-degenerate 4-point relations}
 
In this section we address the Hirota-Miwa relations obeying conditions
A and B given in the beginning of the previous section. Namely, we will 
start with a closer look at non-degenerate 4-point relations, and give
their classification. After that we will show how to obtain various 2-point 
relations by degeneration of the 4-point ones. As we shall see,
many different ways of
degeneration are possible, and that is why 
the classification of 2-point relations is
much more complicated than that of the (non-degenerate) 4-point 
ones\footnote{We do not consider 4-point relations that can be
obtained as degenerations of $M$-point ones for $M>4$ because they 
contain more than four terms.}.

First of all, consider the simplest case of $M=2$ which turns out to be
trivial. Indeed, we should require that the conditions
$$
L^+ + L^- +R^+ + R^- =2, \qquad 
|L^+ - L^-| - |R^+ - R^-|\in 2\ZZ 
$$
hold simultaneously. All possible solutions can be easily found, and it can be
checked that each non-degenerate $2$-point relation 
contains just 2 terms
and converts into identity. Next, for 3-point non-degenerate relations
we have the conditions
$$
L^+ + L^- +R^+ + R^- =3, \qquad 
|L^+ - L^-| - |R^+ - R^-|\in 2\ZZ 
$$
which obviously 
have no common solutions. This means that non-degenerate 3-point 
relations do not exist, and 
any 2- or 3-point Hirota-Miwa relation containing not more than four
terms is a reduction of some 4-point relation.

The case of our main interest is $M=4$ when the conditions are
\beq\label{nd1}
L^+ + L^- +R^+ + R^- =4, \qquad 
|L^+ - L^-| - |R^+ - R^-|\in 2\ZZ .
\eeq
There are many possible solutions. 
However, the number of essentially
different cases can be significantly reduced by taking into account the 
symmetries (\ref{sym1}), (\ref{sym2}), i.e.,
$L^+ \leftrightarrow L^-, R^+ \leftrightarrow R^-$ (simultaneously) and
$(L^+ , L^-) \leftrightarrow (R^+, R^-)$.
First of all we note that the solutions
$$
(L^ +, L^- | R^ +, R^- )=\Bigl \{ (4,0|0,0), (3,1|0,0), (2,2|0,0)\Bigr \}
$$
and
$$
(L^ +, L^- | R^ +, R^- )=\Bigl \{ (0,0|4,0), (0,0|3,1), (0,0|2,2)\Bigr \}
$$
correspond to only one of the DKP subhierarchies (left or right), i.e. the
cases when either right- or left-hand side of equation (\ref{main1}) 
vanishes identically. The multi-component DKP hierarchies were considered
in detail in \cite{SZ24}, so we will not discuss these cases here.
Of our prime interest here are ``intertwining'' relations which mix the
two DKP copies, i.e. solutions to (\ref{nd1}) in which $R^+, R^-$
(or $L^+, L^-$) are not equal to zero simultaneously.
Taking into account the symmetries, we have the following 8 
solutions\footnote{As we shall see below, some of them actually lead to equivalent
equations, so the total number of independent relations is in fact 4.}:
\beq\label{nd2}
\begin{array}{lll}
(L^ +, L^- | R^ +, R^- )&=&\Bigl \{ (3,0|1,0), (3,0|0,1), (2,1|1,0),
(2,1|0,1), 
\\ && \\
&& (2,0|2,0), (2,0|0,2), (2,0|1,1), (1,1|1,1)\Bigr \}.
\end{array}
\eeq

Now, let us write down the equations explicitly. For the notational simplicity 
we use the notation $a,b,c,d\in \CC$ for the four points and 
$\alpha , \beta , \nu , \mu \in \{1, \ldots , N\}$ for the four 
corresponding indices\footnote{Each index is ``linked'' to the corresponding
point: $\alpha$ to $a$, $\beta$ to $b$, $\nu$ to $c$, $\mu$ to $d$.}.
Actually, some of them may coincide but we start with the general 
non-degenerate case
assuming that all of them are distinct. Possible degenerations deserve 
a special consideration and will be
addressed later.

For the first possibility, 
$(L^ +, L^- | R^ +, R^- )=(3,0|1,0)$, we have:
\beq\label{3010}
\begin{array}{rl}
{\bf n}-{\bf n}'={\bf e}_{\alpha}+{\bf e}_{\beta}+{\bf e}_{\nu}, 
\quad & \bar {\bf n}-\bar {\bf n}'={\bf e}_{\mu},
\\ & \\
{\bf t}-{\bf t}'=[a^{-1}]_{\alpha} +[b^{-1}]_{\beta}+ [c^{-1}]_{\nu},
\quad & \bar {\bf t}-\bar {\bf t}'=[d^{-1}]_{\mu}.
\end{array}
\eeq
Equation (\ref{main}) in this case acquires the form
\beq\label{3010a}
\begin{array}{l}
\displaystyle{
\sum_{\gamma =1}^N \epsilon_{\alpha \gamma}\epsilon_{\beta \gamma}
\epsilon_{\nu \gamma}\oint_{C_{\infty}} \left [
\Bigl (\frac{1}{z^{-1}-a^{-1}}\Bigr )^{\delta_{\alpha \gamma}}
\Bigl (\frac{1}{z^{-1}-b^{-1}}\Bigr )^{\delta_{\beta \gamma}}
\Bigl (\frac{1}{z^{-1}-c^{-1}}\Bigr )^{\delta_{\nu \gamma}}
\right. }
\\ \\
\displaystyle{\left. \phantom{aaaaaaaaaaaaaaaaaaa}
\vphantom{\Bigl (\frac{1}{z^{-1}-a^{-1}}\Bigr )^{\delta_{\alpha \gamma}}}
\times \, \tau \Bigl ({\bf n}-{\bf e}_{\gamma}, \bar {\bf n}, {\bf t}-
[z^{-1}]_{\gamma} , \bar {\bf t}\Bigr )
\tau \Bigl ({\bf n}'+{\bf e}_{\gamma}, \bar {\bf n}', {\bf t}+
[z^{-1}]_{\gamma} , \bar {\bf t}'\Bigr )\right ]
}
\\ \\
\displaystyle{
=\, \sum_{\gamma =1}^N \epsilon_{\mu \gamma}
\Bigl (\frac{1}{z^{-1}-d^{-1}}\Bigr )^{\delta_{\mu \gamma}}
\tau \Bigl ({\bf n}, \bar {\bf n}-{\bf e}_{\gamma}, {\bf t},
\bar {\bf t}-[z^{-1}]_{\gamma} ,\Bigr )
\tau \Bigl ({\bf n}', \bar {\bf n}'+{\bf e}_{\gamma}, {\bf t}',
\bar {\bf t}'+[z^{-1}]_{\gamma} ,\Bigr )},
\end{array}
\eeq
where we have written down only non-zero terms. The residue calculus yields
the following Hirota-Miwa equation:
\beq\label{3010b}
\begin{array}{l}
\displaystyle{
\epsilon_{\beta \alpha}\epsilon_{\nu \alpha}
\Bigl (\frac{1}{a^{-1}-b^{-1}}\Bigr )^{\delta_{\beta \alpha}}
\Bigl (\frac{1}{a^{-1}-c^{-1}}\Bigr )^{\delta_{\nu \alpha}}
\tau \Bigl ({\bf n}-{\bf e}_{\alpha}, \bar {\bf n}, {\bf t}-
[a^{-1}]_{\alpha} , \bar {\bf t}\Bigr )}
\\ \\
\displaystyle{\phantom{aaaaaaaaaaaaaaaa}
\times \tau \Bigl ({\bf n}-{\bf e}_{\beta}-{\bf e}_{\nu},
\bar {\bf n}-{\bf e}_{\mu}, {\bf t}-
[b^{-1}]_{\beta}-[c^{-1}]_{\nu} , \bar {\bf t}-[d^{-1}]_{\mu}\Bigr )
}
\\ \\
\displaystyle{
+ \, \epsilon_{\nu \beta}\epsilon_{\alpha \beta}
\Bigl (\frac{1}{b^{-1}-c^{-1}}\Bigr )^{\delta_{\nu \beta}}
\Bigl (\frac{1}{b^{-1}-a^{-1}}\Bigr )^{\delta_{\alpha \beta}}
\tau \Bigl ({\bf n}-{\bf e}_{\beta}, \bar {\bf n}, {\bf t}-
[b^{-1}]_{\beta} , \bar {\bf t}\Bigr )}
\end{array}
\eeq
$$
\begin{array}{l}
\displaystyle{\phantom{aaaaaaaaaaaaaaaa}
\times \tau \Bigl ({\bf n}-{\bf e}_{\nu}-{\bf e}_{\alpha},
\bar {\bf n}-{\bf e}_{\mu}, {\bf t}-
[c^{-1}]_{\nu}-[a^{-1}]_{\alpha} , \bar {\bf t}-[d^{-1}]_{\mu}\Bigr )
}
\\ \\
\displaystyle{
+ \, \epsilon_{\alpha \nu}\epsilon_{\beta \nu}
\Bigl (\frac{1}{c^{-1}-a^{-1}}\Bigr )^{\delta_{\alpha \nu}}
\Bigl (\frac{1}{c^{-1}-b^{-1}}\Bigr )^{\delta_{\beta \nu}}
\tau \Bigl ({\bf n}-{\bf e}_{\nu}, \bar {\bf n}, {\bf t}-
[c^{-1}]_{\nu} , \bar {\bf t}\Bigr )}
\\ \\
\displaystyle{\phantom{aaaaaaaaaaaaaaaa}
\times \tau \Bigl ({\bf n}-{\bf e}_{\alpha}-{\bf e}_{\beta},
\bar {\bf n}-{\bf e}_{\mu}, {\bf t}-
[a^{-1}]_{\alpha}-[b^{-1}]_{\beta} , \bar {\bf t}-[d^{-1}]_{\mu}\Bigr )
}
\\ \\
=\, \tau \Bigl ({\bf n},
\bar {\bf n}-{\bf e}_{\mu}, {\bf t}, \bar {\bf t}-[d^{-1}]_{\mu}\Bigr )
\tau \Bigl ({\bf n}-{\bf e}_{\alpha}-{\bf e}_{\beta} -{\bf e}_{\nu},
\bar {\bf n}, {\bf t}-
[a^{-1}]_{\alpha}-[b^{-1}]_{\beta}-[c^{-1}]_{\nu} , 
\bar {\bf t}\Bigr ),
\end{array}
$$
or, using the notation (\ref{m5}),
\beq\label{3010c}
\begin{array}{l}
\displaystyle{
E^{-1}_{\beta \alpha}(a,b)E^{-1}_{\nu \alpha}(a,c)
\tau \Bigl ({\bf n}-{\bf e}_{\alpha}, \bar {\bf n}, {\bf t}-
[a^{-1}]_{\alpha} , \bar {\bf t}\Bigr )}
\\ \\
\displaystyle{\phantom{aaaaaaaaaaaaaaaa}
\times \tau \Bigl ({\bf n}-{\bf e}_{\beta}-{\bf e}_{\nu},
\bar {\bf n}-{\bf e}_{\mu}, {\bf t}-
[b^{-1}]_{\beta}-[c^{-1}]_{\nu} , \bar {\bf t}-[d^{-1}]_{\mu}\Bigr )
}
\\ \\
\displaystyle{
+ \, E^{-1}_{\nu \beta}(b,c)E^{-1}_{\alpha \beta}(b,a)
\tau \Bigl ({\bf n}-{\bf e}_{\beta}, \bar {\bf n}, {\bf t}-
[b^{-1}]_{\beta} , \bar {\bf t}\Bigr )}
\\ \\
\displaystyle{\phantom{aaaaaaaaaaaaaaaa}
\times \tau \Bigl ({\bf n}-{\bf e}_{\nu}-{\bf e}_{\alpha},
\bar {\bf n}-{\bf e}_{\mu}, {\bf t}-
[c^{-1}]_{\nu}-[a^{-1}]_{\alpha} , \bar {\bf t}-[d^{-1}]_{\mu}\Bigr )
}
\\ \\
\displaystyle{
+ \, E^{-1}_{\alpha \nu}(c,a)E^{-1}_{\beta \nu}(c,b)
\epsilon_{\alpha \nu}\epsilon_{\beta \nu}
\tau \Bigl ({\bf n}-{\bf e}_{\nu}, \bar {\bf n}, {\bf t}-
[c^{-1}]_{\nu} , \bar {\bf t}\Bigr )}
\\ \\
\displaystyle{\phantom{aaaaaaaaaaaaaaaa}
\times \tau \Bigl ({\bf n}-{\bf e}_{\alpha}-{\bf e}_{\beta},
\bar {\bf n}-{\bf e}_{\mu}, {\bf t}-
[a^{-1}]_{\alpha}-[b^{-1}]_{\beta} , \bar {\bf t}-[d^{-1}]_{\mu}\Bigr )
}
\\ \\
=\, \tau \Bigl ({\bf n},
\bar {\bf n}-{\bf e}_{\mu}, {\bf t}, \bar {\bf t}-[d^{-1}]_{\mu}\Bigr )
\tau \Bigl ({\bf n}-{\bf e}_{\alpha}-{\bf e}_{\beta} -{\bf e}_{\nu},
\bar {\bf n}, {\bf t}-
[a^{-1}]_{\alpha}-[b^{-1}]_{\beta}-[c^{-1}]_{\nu} , 
\bar {\bf t}\Bigr ).
\end{array}
\eeq

For the second possibility in (\ref{nd2}), 
$(L^ +, L^- | R^ +, R^- )=(3,0|0,1)$, we have:
\beq\label{3001}
\begin{array}{rl}
{\bf n}-{\bf n}'={\bf e}_{\alpha}+{\bf e}_{\beta}+{\bf e}_{\nu}, 
\quad & \bar {\bf n}-\bar {\bf n}'=-{\bf e}_{\mu},
\\ & \\
{\bf t}-{\bf t}'=[a^{-1}]_{\alpha} +[b^{-1}]_{\beta}+ [c^{-1}]_{\nu},
\quad & \bar {\bf t}-\bar {\bf t}'=-[d^{-1}]_{\mu},
\end{array}
\eeq
which leads to the following Hirota-Miwa equation:
\beq\label{3001a}
\begin{array}{l}
\displaystyle{
E^{-1}_{\beta \alpha}(a,b)E^{-1}_{\nu \alpha}(a,c)
\tau \Bigl ({\bf n}-{\bf e}_{\alpha}, \bar {\bf n}, {\bf t}-
[a^{-1}]_{\alpha} , \bar {\bf t}\Bigr )}
\\ \\
\displaystyle{\phantom{aaaaaaaaaaaaaaaa}
\times \tau \Bigl ({\bf n}-{\bf e}_{\beta}-{\bf e}_{\nu},
\bar {\bf n}+{\bf e}_{\mu}, {\bf t}-
[b^{-1}]_{\beta}-[c^{-1}]_{\nu} , \bar {\bf t}+[d^{-1}]_{\mu}\Bigr )
}
\\ \\
\displaystyle{
+ \, E^{-1}_{\nu \beta}(b,c)E^{-1}_{\alpha \beta}(b,a)
\tau \Bigl ({\bf n}-{\bf e}_{\beta}, \bar {\bf n}, {\bf t}-
[b^{-1}]_{\beta} , \bar {\bf t}\Bigr )}
\\ \\
\displaystyle{\phantom{aaaaaaaaaaaaaaaa}
\times \tau \Bigl ({\bf n}-{\bf e}_{\nu}-{\bf e}_{\alpha},
\bar {\bf n}+{\bf e}_{\mu}, {\bf t}-
[c^{-1}]_{\nu}-[a^{-1}]_{\alpha} , \bar {\bf t}+[d^{-1}]_{\mu}\Bigr )
}
\end{array}
\eeq
$$
\begin{array}{l}
\displaystyle{
+ \, E^{-1}_{\alpha \nu}(c,a)E^{-1}_{\beta \nu}(c,b)
\tau \Bigl ({\bf n}-{\bf e}_{\nu}, \bar {\bf n}, {\bf t}-
[c^{-1}]_{\nu} , \bar {\bf t}\Bigr )}
\\ \\
\displaystyle{\phantom{aaaaaaaaaaaaaaaa}
\times \tau \Bigl ({\bf n}-{\bf e}_{\alpha}-{\bf e}_{\beta},
\bar {\bf n}+{\bf e}_{\mu}, {\bf t}-
[a^{-1}]_{\alpha}-[b^{-1}]_{\beta} , \bar {\bf t}+[d^{-1}]_{\mu}\Bigr )
}
\\ \\
=\, \tau \Bigl ({\bf n},
\bar {\bf n}+{\bf e}_{\mu}, {\bf t}, \bar {\bf t}+[d^{-1}]_{\mu}\Bigr )
\tau \Bigl ({\bf n}-{\bf e}_{\alpha}-{\bf e}_{\beta} -{\bf e}_{\nu},
\bar {\bf n}, {\bf t}-
[a^{-1}]_{\alpha}-[b^{-1}]_{\beta}-[c^{-1}]_{\nu} , 
\bar {\bf t}\Bigr ).
\end{array}
$$

For the third possibility in (\ref{nd2}), 
$(L^ +, L^- | R^ +, R^- )=(2,1|1,0)$, we have:
\beq\label{2110}
\begin{array}{rl}
{\bf n}-{\bf n}'={\bf e}_{\alpha}+{\bf e}_{\beta}-{\bf e}_{\nu}, 
\quad & \bar {\bf n}-\bar {\bf n}'={\bf e}_{\mu},
\\ & \\
{\bf t}-{\bf t}'=[a^{-1}]_{\alpha} +[b^{-1}]_{\beta}- [c^{-1}]_{\nu},
\quad & \bar {\bf t}-\bar {\bf t}'=[d^{-1}]_{\mu},
\end{array}
\eeq
which leads to the equation
\beq\label{2110a}
\begin{array}{l}
\displaystyle{
E^{-1}_{\alpha \beta}(b,a)E_{\nu \beta}(b,c)
\tau \Bigl ({\bf n}-{\bf e}_{\alpha}+{\bf e}_{\nu}, \bar {\bf n}
-{\bf e}_{\mu}, {\bf t}-
[a^{-1}]_{\alpha} +[c^{-1}]_{\nu}, \bar {\bf t}-[d^{-1}]_{\mu}\Bigr )}
\\ \\
\displaystyle{\phantom{aaaaaaaaaaaaaaaa}
\times \tau \Bigl ({\bf n}-{\bf e}_{\beta},
\bar {\bf n}, {\bf t}-
[b^{-1}]_{\beta}, \bar {\bf t}\Bigr )
}
\\ \\
\displaystyle{
+ \, E^{-1}_{ \beta \alpha}(a,b)E_{\nu \alpha}(a,c)
\tau \Bigl ({\bf n}-{\bf e}_{\beta}+{\bf e}_{\nu}, 
\bar {\bf n}-{\bf e}_{\mu}, {\bf t}-
[b^{-1}]_{\beta}+[c^{-1}]_{\nu}  , \bar {\bf t}-[d^{-1}]_{\mu} \Bigr )}
\\ \\
\displaystyle{\phantom{aaaaaaaaaaaaaaaa}
\times \tau \Bigl ({\bf n}-{\bf e}_{\alpha},
\bar {\bf n}, {\bf t}-
[a^{-1}]_{\alpha}, \bar {\bf t}\Bigr )
}
\\ \\
\displaystyle{
+ \, E_{\alpha \nu}(c,a)E_{\beta \nu}(c,b)
\tau \Bigl ({\bf n}+{\bf e}_{\nu}, \bar {\bf n}, {\bf t}+
[c^{-1}]_{\nu} , \bar {\bf t}\Bigr )}
\\ \\
\displaystyle{\phantom{aaaaaaaaaaaaaaaa}
\times \tau \Bigl ({\bf n}-{\bf e}_{\alpha}-{\bf e}_{\beta},
\bar {\bf n}-{\bf e}_{\mu}, {\bf t}-
[a^{-1}]_{\alpha}-[b^{-1}]_{\beta} , \bar {\bf t}-[d^{-1}]_{\mu}\Bigr )
}
\\ \\
=\, \tau \Bigl ({\bf n},
\bar {\bf n}-{\bf e}_{\mu}, {\bf t}, \bar {\bf t}-[d^{-1}]_{\mu}\Bigr )
\tau \Bigl ({\bf n}-{\bf e}_{\alpha}-{\bf e}_{\beta} +{\bf e}_{\nu},
\bar {\bf n}, {\bf t}-
[a^{-1}]_{\alpha}-[b^{-1}]_{\beta}+[c^{-1}]_{\nu} , 
\bar {\bf t}\Bigr ).
\end{array}
\eeq
It is easy to see that this equation is in fact 
the same as (\ref{3001a}). Indeed,
multiplying all terms of the latter by $E_{\alpha \nu}(c,a)E_{\beta \nu}(c,b)$,
shifting the variables as
$$
\begin{array}{l}
({\bf n}, {\bf t}) \longrightarrow ({\bf n}+{\bf e}_{\nu}, {\bf t}+
[c^{-1}]_{\nu}),
\\ \\
(\bar {\bf n}, \bar {\bf t}) 
\longrightarrow (\bar {\bf n} -{\bf e}_{\mu}, \bar {\bf t}-[d^{-1}]_{\mu})
\end{array}
$$
and rearranging terms, we get equation (\ref{2110a}).
In the same way, one can see that the equation obtained for the choice
$(L^ +, L^- | R^ +, R^- )=(2,1|0,1)$ is the same as equation (\ref{3010c}).

Next, consider the case
$(L^ +, L^- | R^ +, R^- )=(2,0|2,0)$, i.e.
\beq\label{2020}
\begin{array}{rl}
{\bf n}-{\bf n}'={\bf e}_{\alpha}+{\bf e}_{\beta}, 
\quad & \bar {\bf n}-\bar {\bf n}'={\bf e}_{\nu}+{\bf e}_{\mu},
\\ & \\
{\bf t}-{\bf t}'=[a^{-1}]_{\alpha} +[b^{-1}]_{\beta},
\quad & \bar {\bf t}-\bar {\bf t}'=[c^{-1}]_{\nu}+[d^{-1}]_{\mu},
\end{array}
\eeq
which leads to the equation
\beq\label{2020a}
\begin{array}{l}
\displaystyle{
E^{-1}_{\beta \alpha}(a,b)
\tau \Bigl ({\bf n}-{\bf e}_{\alpha}, \bar {\bf n}, {\bf t}-
[a^{-1}]_{\alpha}, \bar {\bf t}\Bigr )}
\\ \\
\displaystyle{\phantom{aaaaaaaaaaaa}
\times \tau \Bigl ({\bf n}-{\bf e}_{\beta},
\bar {\bf n}-{\bf e}_{\nu}-{\bf e}_{\mu}, {\bf t}-
[b^{-1}]_{\beta}, \bar {\bf t}-[c^{-1}]_{\nu}-[d^{-1}]_{\mu}\Bigr )
}
\\ \\
\displaystyle{
+ \, E^{-1}_{ \alpha \beta }(b,a)
\tau \Bigl ({\bf n}-{\bf e}_{\beta}, 
\bar {\bf n}, {\bf t}-
[b^{-1}]_{\beta} , \bar {\bf t} \Bigr )}
\\ \\
\displaystyle{\phantom{aaaaaaaaaaaa}
\times \tau \Bigl ({\bf n}-{\bf e}_{\alpha},
\bar {\bf n}-{\bf e}_{\nu}-{\bf e}_{\mu}, {\bf t}-
[a^{-1}]_{\alpha}, \bar {\bf t}-[c^{-1}]_{\nu}-[d^{-1}]_{\mu}\Bigr )
}
\end{array}
\eeq
$$
\begin{array}{l}
\displaystyle{
= \, E^{-1}_{\mu \nu}(c,d)
\tau \Bigl ({\bf n}, \bar {\bf n}-{\bf e}_{\nu}, {\bf t},
\bar {\bf t}-[c^{-1}]_{\nu}\Bigr )}
\\ \\
\displaystyle{\phantom{aaaaaaaaaaaa}
\times \tau \Bigl ({\bf n}-{\bf e}_{\alpha}-{\bf e}_{\beta},
\bar {\bf n}-{\bf e}_{\mu}, {\bf t}-
[a^{-1}]_{\alpha}-[b^{-1}]_{\beta} , \bar {\bf t}-[d^{-1}]_{\mu}\Bigr )}
\\ \\
+\, E^{-1}_{\nu \mu}(d,c)
\tau \Bigl ({\bf n},
\bar {\bf n}-{\bf e}_{\mu}, {\bf t}, \bar {\bf t}-[d^{-1}]_{\mu}\Bigr )
\\ \\
\phantom{aaaaaaaaaaa}\times 
\tau \Bigl ({\bf n}-{\bf e}_{\alpha}-{\bf e}_{\beta},
\bar {\bf n}-{\bf e}_{\nu}, {\bf t}-
[a^{-1}]_{\alpha}-[b^{-1}]_{\beta} , 
\bar {\bf t}-[c^{-1}]_{\nu}\Bigr ).
\end{array}
$$
By the procedure similar to the one explained above, one can show
that the equations for 
$(L^ +, L^- | R^ +, R^- )=(2,0|0,2)$ and
$(L^ +, L^- | R^ +, R^- )=(1,1|1,1)$ are equivalent to (\ref{2020a}).

At last, consider the remaining case
$(L^ +, L^- | R^ +, R^- )=(2,0|1,1)$, i.e.
\beq\label{2011}
\begin{array}{rl}
{\bf n}-{\bf n}'={\bf e}_{\alpha}+{\bf e}_{\beta}, 
\quad & \bar {\bf n}-\bar {\bf n}'={\bf e}_{\nu}-{\bf e}_{\mu},
\\ & \\
{\bf t}-{\bf t}'=[a^{-1}]_{\alpha} +[b^{-1}]_{\beta},
\quad & \bar {\bf t}-\bar {\bf t}'=[c^{-1}]_{\nu}-[d^{-1}]_{\mu}.
\end{array}
\eeq
The corresponding Hirota-Miwa equation is as follows:
\beq\label{2011a}
\begin{array}{l}
\displaystyle{
E^{-1}_{\beta \alpha}(a,b)
\tau \Bigl ({\bf n}-{\bf e}_{\alpha}, \bar {\bf n}, {\bf t}-
[a^{-1}]_{\alpha}, \bar {\bf t}\Bigr )}
\\ \\
\displaystyle{\phantom{aaaaaaaaaaaa}
\times \tau \Bigl ({\bf n}-{\bf e}_{\beta},
\bar {\bf n}-{\bf e}_{\nu}+{\bf e}_{\mu}, {\bf t}-
[b^{-1}]_{\beta}, \bar {\bf t}-[c^{-1}]_{\nu}+[d^{-1}]_{\mu}\Bigr )
}
\\ \\
\displaystyle{
+ \, E^{-1}_{ \alpha \beta }(b,a)
\tau \Bigl ({\bf n}-{\bf e}_{\beta}, 
\bar {\bf n}, {\bf t}-
[b^{-1}]_{\beta} , \bar {\bf t} \Bigr )}
\\ \\
\displaystyle{\phantom{aaaaaaaaaaaa}
\times \tau \Bigl ({\bf n}-{\bf e}_{\alpha},
\bar {\bf n}-{\bf e}_{\nu}+{\bf e}_{\mu}, {\bf t}-
[a^{-1}]_{\alpha}, \bar {\bf t}-[c^{-1}]_{\nu}+[d^{-1}]_{\mu}\Bigr )
}
\end{array}
\eeq
$$
\begin{array}{l}
\displaystyle{
= \, E_{\mu \nu}(c,d)
\tau \Bigl ({\bf n}, \bar {\bf n}-{\bf e}_{\nu}, {\bf t},
\bar {\bf t}-[c^{-1}]_{\nu}\Bigr )}
\\ \\
\displaystyle{\phantom{aaaaaaaaaaaa}
\times \tau \Bigl ({\bf n}-{\bf e}_{\alpha}-{\bf e}_{\beta},
\bar {\bf n}+{\bf e}_{\mu}, {\bf t}-
[a^{-1}]_{\alpha}-[b^{-1}]_{\beta} , \bar {\bf t}+[d^{-1}]_{\mu}\Bigr )}
\\ \\
+\, E_{\nu \mu}(d,c)
\tau \Bigl ({\bf n},
\bar {\bf n}+{\bf e}_{\mu}, {\bf t}, \bar {\bf t}+[d^{-1}]_{\mu}\Bigr )
\\ \\
\phantom{aaaaaaaaaaa}\times 
\tau \Bigl ({\bf n}-{\bf e}_{\alpha}-{\bf e}_{\beta},
\bar {\bf n}-{\bf e}_{\nu}, {\bf t}-
[a^{-1}]_{\alpha}-[b^{-1}]_{\beta} , 
\bar {\bf t}-[c^{-1}]_{\nu}\Bigr ).
\end{array}
$$

Summing up, there are
four independent 4-point ``intertwining'' relations, corresponding to
\beq\label{nd5}
(L^ +, L^- | R^ +, R^- )=
\Bigl \{ (3,0|1,0), (3,0|0,1), (2,0|2,0), (2,0|1,1)\Bigr \}.
\eeq
They are (\ref{3010c}), (\ref{3001a}), (\ref{2020a}), (\ref{2011a}), respectively.
Other relations can be obtained from them by applying the apparent symmetries
(\ref{sym1}), (\ref{sym2}). As we have observed, in some cases different 
solutions from
the list (\ref{nd2}) not connected by any of the two symmetry
transformations (for example, $(3,0|1,0)$ and $(2,1|0,1)$)
yield identical equations. This fact is probably due to some hidden 
symmetry of equation (\ref{main}).

At last, let us rewrite the 4-term relations in another form,
suitable for subsequent degeneration to 2-point ones.
In order to represent the equations in a more compact form,
the following short-hand notation for the independent
discrete variables are useful:
\beq\label{nd6}
{\bf n}^{\alpha}={\bf n} +{\bf e}_{\alpha}, 
\quad
{\bf n}_{\alpha}={\bf n} -{\bf e}_{\alpha},
\quad
{\bf n}^{\alpha \beta}={\bf n} +{\bf e}_{\alpha}+{\bf e}_{\beta}
\quad
{\bf n}^{\alpha}_{\beta}={\bf n} +{\bf e}_{\alpha}-{\bf e}_{\beta}, 
\eeq
and so on,
and similarly for $\bar {\bf n}$. For example, 
${\bf n}^{\alpha \beta}_{\nu}={\bf n} +{\bf e}_{\alpha}+{\bf e}_{\beta}
-{\bf e}_{\nu}$. Similar notations will be used for shifts of continuous
times:
\beq\label{nd6a}
{\bf t}^{[a_{\alpha}]}={\bf t} + [a^{-1}]_{\alpha}, 
\quad
{\bf t}^{[a_{\alpha}b_{\beta}]}={\bf t} + [a^{-1}]_{\alpha}+ [b^{-1}]_{\beta},
\quad
{\bf t}^{[a_{\alpha}]}_{[b_{\beta}]}={\bf t} + [a^{-1}]_{\alpha}- [b^{-1}]_{\beta},
\eeq
and so on, and similarly for the times $\bar {\bf t}$.
For example,
the tau-function $\tau ({\bf n}+{\bf e}_{\alpha}, \bar {\bf n}-
{\bf e}_{\nu}, {\bf t}+[a^{-1}]_{\alpha}+ [b^{-1}]_{\beta}, \bar {\bf t})$
will be written as
$$
\tau ({\bf n}+{\bf e}_{\alpha}, \bar {\bf n}-
{\bf e}_{\nu}, {\bf t}+[a^{-1}]_{\alpha}+ [b^{-1}]_{\beta}, \bar {\bf t})=
\tau ({\bf n}^{\alpha}, \bar {\bf n}_{\beta}, {\bf t}^{[a_{\alpha}b_{\beta}]},
\bar {\bf t}).
$$
In this notation, the four basic
equations for $(L^ +, L^- | R^ +, R^- )$ as in (\ref{nd5})
acquire the following form (after some shifts of the variables):

\begin{itemize}
\item
For $(3,0|1,0)$:
\beq\label{3010-compact}
\begin{array}{ll}
&\epsilon_{\beta \nu} (c^{-1}-b^{-1})^{\delta_{\beta \nu}}
\tau \Bigl ({\bf n}^{\beta \nu}, \bar {\bf n}^{\mu}, 
{\bf t}^{[b_{\beta}c_{\nu}]}, \bar {\bf t}^{[d_{\mu}]}\Bigr )
\tau \Bigl ({\bf n}^{\alpha}, \bar {\bf n}, {\bf t}^{[a_{\alpha}]}, \bar {\bf t}
\Bigr )
\\ & \\
+ & \hspace{-2mm}
\epsilon_{ \nu \alpha} (a^{-1}-c^{-1})^{\delta_{\nu \alpha}}
\tau \Bigl ({\bf n}^{\nu \alpha}, \bar {\bf n}^{\mu}, 
{\bf t}^{[c_{\nu}a_{\alpha}]}, \bar {\bf t}^{[d_{\mu}]}\Bigr )
\tau \Bigl ({\bf n}^{\beta}, \bar {\bf n}, {\bf t}^{[b_{\beta}]}, \bar {\bf t}
\Bigr )
\\ & \\
+ & \hspace{-2mm}
\epsilon_{\alpha \beta} (b^{-1}-a^{-1})^{\delta_{\alpha \beta}}
\tau \Bigl ({\bf n}^{\alpha \beta}, \bar {\bf n}^{\mu}, 
{\bf t}^{[a_{\alpha}b_{\beta}]}, \bar {\bf t}^{[d_{\mu}]}\Bigr )
\tau \Bigl ({\bf n}^{\nu}, \bar {\bf n}, {\bf t}^{[c_{\nu}]}, \bar {\bf t}
\Bigr )
\\ & \\
+ & \hspace{-2mm}
\epsilon_{\alpha \beta \nu}
(b^{-1}-a^{-1})^{\delta_{\alpha \beta}}
(c^{-1}-b^{-1})^{\delta_{\beta \nu}}
(a^{-1}-c^{-1})^{\delta_{\nu \alpha}}
\\ & \\
& \phantom{aaaaaaaaaaaaaaa}
\times \, \tau \Bigl ({\bf n}^{\alpha \beta \nu}, \bar {\bf n}, 
{\bf t}^{[a_{\alpha}b_{\beta}c_{\nu}]}, \bar {\bf t}\Bigr )
\tau \Bigl ({\bf n}, \bar {\bf n}^{\mu}, {\bf t}, \bar {\bf t}^{[d_{\mu}]}\Bigr )
=0,
\end{array}
\eeq
where $\epsilon_{\alpha \beta \nu} =\epsilon_{\alpha \beta}\epsilon_{ \beta \nu}
\epsilon_{\nu \alpha}$;

\item
For $(3,0|0,1)$:
\beq\label{3001-compact}
\begin{array}{ll}
&\epsilon_{\beta \nu} (c^{-1}-b^{-1})^{\delta_{\beta \nu}}
\tau \Bigl ({\bf n}^{\beta \nu}, \bar {\bf n}, 
{\bf t}^{[b_{\beta}c_{\nu}]}, \bar {\bf t}\Bigr )
\tau \Bigl ({\bf n}^{\alpha}, \bar {\bf n}^{\mu}, 
{\bf t}^{[a_{\alpha}]}, \bar {\bf t}^{[d_{\mu}]}
\Bigr )
\\ & \\
+ & \hspace{-2mm}
\epsilon_{ \nu \alpha} (a^{-1}-c^{-1})^{\delta_{\nu \alpha}}
\tau \Bigl ({\bf n}^{\nu \alpha}, \bar {\bf n}, 
{\bf t}^{[c_{\nu}a_{\alpha}]}, \bar {\bf t}\Bigr )
\tau \Bigl ({\bf n}^{\beta}, \bar {\bf n}^{\mu}, {\bf t}^{[b_{\beta}]}, 
\bar {\bf t}^{[d_{\mu}]}
\Bigr )
\\ & \\
+ & \hspace{-2mm}
\epsilon_{\alpha \beta} (b^{-1}-a^{-1})^{\delta_{\alpha \beta}}
\tau \Bigl ({\bf n}^{\alpha \beta}, \bar {\bf n}, 
{\bf t}^{[a_{\alpha}b_{\beta}]}, \bar {\bf t}\Bigr )
\tau \Bigl ({\bf n}^{\nu}, \bar {\bf n}^{\mu}, {\bf t}^{[c_{\nu}]}, 
\bar {\bf t}^{[d_{\mu}]}
\Bigr )
\\ & \\
+ & \hspace{-2mm}
\epsilon_{\alpha \beta \nu}
(b^{-1}-a^{-1})^{\delta_{\alpha \beta}}
(c^{-1}-b^{-1})^{\delta_{\beta \nu}}
(a^{-1}-c^{-1})^{\delta_{\nu \alpha}}
\\ & \\
& \phantom{aaaaaaaaaaaaaaa}
\times \, \tau \Bigl ({\bf n}^{\alpha \beta \nu}, \bar {\bf n}^{\mu}, 
{\bf t}^{[a_{\alpha}b_{\beta}c_{\nu}]}, \bar {\bf t}^{[d_{\mu}]}\Bigr )
\tau \Bigl ({\bf n}, \bar {\bf n}, {\bf t}, \bar {\bf t}\Bigr )
=0
\end{array}
\eeq

\item
For $(2,0|2,0)$:
\beq\label{2020-compact}
\begin{array}{ll}
\displaystyle{
\epsilon_{\mu \nu}(c^{-1}-d^{-1})^{\delta_{\nu \mu}}
\left [ \vphantom{\frac{a}{b}}
\tau \Bigl ({\bf n}^{\beta}, \bar {\bf n}^{\nu \mu}, 
{\bf t}^{[b_{\beta}]}, \bar {\bf t}^{[c_{\nu}d_{\mu}]}\Bigr )
\tau \Bigl ({\bf n}^{\alpha}, \bar {\bf n}, 
{\bf t}^{[a_{\alpha}]}, \bar {\bf t}\Bigr )\right.}
\\ \\
\displaystyle{
\left. \vphantom{\frac{a}{b}} \phantom{aaaaaaaaaaaaaaaaaaa}
-\, \tau \Bigl ({\bf n}^{\alpha}, \bar {\bf n}^{\nu \mu}, 
{\bf t}^{[a_{\alpha}]}, \bar {\bf t}^{[c_{\nu}d_{\mu}]}\Bigr )
\tau \Bigl ({\bf n}^{\beta}, \bar {\bf n}, 
{\bf t}^{[b_{\beta}]}, \bar {\bf t}\Bigr ) \right ]}
\\ \\
\displaystyle{
=\, \epsilon_{\beta \alpha}(a^{-1}-b^{-1})^{\delta_{\alpha \beta}}
\left [ \vphantom{\frac{a}{b}}
\tau \Bigl ({\bf n}^{\alpha \beta}, \bar {\bf n}^{\mu}, 
{\bf t}^{[a_{\alpha}b_{\beta}]}, \bar {\bf t}^{[d_{\mu}]}\Bigr )
\tau \Bigl ({\bf n}, \bar {\bf n}^{\nu}, 
{\bf t}, \bar {\bf t}^{[c_{\nu}]}\Bigr )\right.}
\\ \\
\displaystyle{
\left. \vphantom{\frac{a}{b}} \phantom{aaaaaaaaaaaaaaaaaaa}
-\, \tau \Bigl ({\bf n}^{\alpha \beta}, \bar {\bf n}^{\nu}, 
{\bf t}^{[a_{\alpha}b_{\beta}]}, \bar {\bf t}^{[c_{\nu}d_{\mu}]}\Bigr )
\tau \Bigl ({\bf n}^{\beta}, \bar {\bf n}, 
{\bf t}^{[a_{\alpha}b_{\beta}]}, \bar {\bf t}^{[c_{\nu}]}
\Bigr ) \right ]};
\end{array}
\eeq

\item
For $(2,0|1,1)$:
\beq\label{2011-compact}
\begin{array}{ll}
\tau \Bigl ({\bf n}^{\beta}, \bar {\bf n}^{\nu}, 
{\bf t}^{[b_{\beta}]}, \bar {\bf t}^{[c_{\nu}]}\Bigr )
\tau \Bigl ({\bf n}^{\alpha}, \bar {\bf n}^{\mu}, 
{\bf t}^{[a_{\alpha}]}, \bar {\bf t}^{[d_{\mu}]}\Bigr )
\\ \\
\displaystyle{
\phantom{aaaaaaaaaaaaaaaaaaa}
-\, \tau \Bigl ({\bf n}^{\alpha}, \bar {\bf n}^{\nu}, 
{\bf t}^{[a_{\alpha}]}, \bar {\bf t}^{[c_{\nu}]}\Bigr )
\tau \Bigl ({\bf n}^{\beta}, \bar {\bf n}^{\mu}, 
{\bf t}^{[b_{\beta}]}, \bar {\bf t}^{[d_{\mu}]}\Bigr ) }
\\ \\
\displaystyle{
=\, \epsilon_{\beta \alpha}\epsilon_{\mu \nu}
(a^{-1}-b^{-1})^{\delta_{\alpha \beta}}
(c^{-1}-d^{-1})^{\delta_{\nu \mu}} \left [
\vphantom{\frac{a}{b}}
\tau \Bigl ({\bf n}^{\alpha \beta}, \bar {\bf n}, 
{\bf t}^{[a_{\alpha}b_{\beta}]}, \bar {\bf t}\Bigr )
\tau \Bigl ({\bf n}, \bar {\bf n}^{\nu \mu}, 
{\bf t}, \bar {\bf t}^{[c_{\nu}d_{\mu}]}\Bigr )\right.}
\\ \\
\displaystyle{
\left. \vphantom{\frac{a}{b}} \phantom{aaaaaaaaaaaaaaaaaaa}
-\, \tau \Bigl ({\bf n}^{\alpha \beta}, \bar {\bf n}^{\nu \mu}, 
{\bf t}^{[a_{\alpha}b_{\beta}]}, \bar {\bf t}^{[c_{\nu}d_{\mu}]}\Bigr )
\tau \Bigl ({\bf n}, \bar {\bf n}, 
{\bf t}, \bar {\bf t}
\Bigr ) \right ]}.
\end{array}
\eeq
\end{itemize}

\noindent
Note that the two equations (\ref{3010-compact}) and
(\ref{3001-compact}) are connected by the transformation
$(\bar {\bf n}, \bar {\bf t})\to (-\bar {\bf n}, -\bar {\bf t})$.
The full set of non-degenerate 4-point relations contains also
3 ``mirrors'' of equations (\ref{3010-compact}), (\ref{3001-compact}),
(\ref{2011-compact}) obtained from them 
by exchange of the variables with and without bar
(i.e., $({\bf n}, {\bf t})\leftrightarrow (\bar {\bf n}, \bar {\bf t})$).
Equation (\ref{2020-compact}) is invariant under this transformation.

Lastly, we note that if any two points coincide together with the corresponding
indices (for example, if $a=b$ and $\alpha =\beta$), these equations 
convert into
identities $0=0$. However, they can be expanded up to the next term 
of the Taylor expansion in $b^{-1}-a^{-1}=\varepsilon \to 0$, and this gives 
already non-trivial relations containing derivatives with respect to the
continuous times. Another source of derivatives is expansion of these
equations around infinity, as some of the points tend to $\infty$.

To obtain all 4-term 2-point relations from the 4-point ones, we should
tend two of the four points $a,b,c,d$ to infinity. To further 
reduce the number of equations which can be obtained in this way,
we add to conditions A and B the following condition C:

\begin{itemize}
\item[C)]
The equations should contain not more than two different indices
$\alpha , \beta$ (in accordance with the number of points).
\end{itemize}

\noindent
However, even with this requirement, there are quite many (namely, 32) 
2-point equations. The full list of them is given in the Appendix.

\section{Conclusion and discussion of further perspectives}

In this paper, we have introduced the multi-component version
of the Pfaff-Toda hierarchy. (The simplest one-component case was 
earlier suggested by Takasaki in \cite{Takasaki09}.)
Our approach is based on the free fermions technique developed by the 
Kyoto school. The multi-component fermions allow one to define 
multi-component generalizations of known integrable hierarchies,
such as KP, Toda, and their Pfaff-versions in a natural way.
This approach is very well adapted for treating the hierarchies 
within the bilinear formalism, when tau-function, defined as the
expectation value of certain fermionic operators, plays the role
of a universal dependent variable. The operator bilinear identity
for fermions implies that the tau-function satisfies a general
integral bilinear relation, which can be called ``generating'' 
because contains all equations of the hierarchy in a compressed
form. 

In this paper, we have defined the tau-function of the multi-component
Pfaff-Toda hierarchy, $\tau ({\bf n}, \bar {\bf n}, {\bf t},  \bar {\bf t})$,
as the expectation value (\ref{f1}) and proved that
it obeys the integral bilinear relation (\ref{main}). Also, we have 
obtained and classified bilinear equations of Hirota-Miwa type 
(containing 4 bilinear terms) that
follow from it.

An important related problem is to figure out what is a minimal set
of the Hirota-Miwa equations that are equivalent to the whole hierarchy.
For one-component KP, modified KP and BKP hierarchies 
the answer is known \cite{Shigyo13}: just the very first nontrivial
Hirota-Miwa relation (two- or three-point depending on the hierarchy)
is already equivalent to the whole hierarchy. The proof suggested
in \cite{Shigyo13} is based on 
determinant (or pfaffian) formulas that express tau-function 
depending on many Miwa variables as determinant (or pfaffian) of a
matrix whose matrix elements are expressed through tau-function
with only two Miwa variables. In their turn, these formulas may serve
as an alternative (but equivalent) formulation of the hierarchy in question.
To the best of our knowledge, nothing in this respect is known for 
multi-component hierarchies and for hierarchies of the Pfaff type,
even in the one-component case. This is an interesting open problem for 
future research.

This paper has been thought of as a first part of two (the second
part is in progress). The second part is going to be on 
the dispersionless limit of the Pfaff-Toda hierarchy introduced here.
The standard reference to the dispersionless limits of integrable
hierarchies is \cite{TT95}.
In order to perform this limit, one should introduce a small
parameter $\hbar$ (which is eventually tending to 0)
and re-scale the times ${\bf t}$, $\bar {\bf t}$
and the variables ${\bf n}$, $\bar {\bf n}$ as
$t_{\alpha , k}\to t_{\alpha , k}/\hbar$, 
$n_{\alpha}\to t_{\alpha , 0}/\hbar$ and similarly 
for the bar-variables. Introduce the function 
$F({\bf t}_0, \bar {\bf t}_0,  {\bf t}, \bar {\bf t}; 
\hbar )$ related to the tau-function of the multi-component
Pfaff-Toda hierarchy by the
formula
\beq\label{d1}
\tau \Bigl (\hbar^{-1}{\bf t}_0 , \hbar^{-1}\bar {\bf t}_0 , 
\hbar^{-1}{\bf t} , \hbar^{-1}\bar {\bf t} \Bigr )=
\exp \left ( \frac{1}{\hbar^2}\,
F({\bf t}_0, \bar {\bf t}_0, {\bf t},\bar {\bf t}; \hbar )\right )
\eeq
and consider the limit  $F=\lim\limits_{\hbar \to 0}
F({\bf t}_0, \bar {\bf t}_0, {\bf t},\bar {\bf t}; \hbar )$ 
(if it exists\footnote{There are several important case when this
limit does exist, for example, for tau-functions that appear as
partition functions of random matrix models. However, for some classes
of tau-functions (for example, for soliton or algebro-geometric
solutions) this limit does not exist.}). 
The function $F$ (sometimes called free energy) substitutes the 
tau-function in the dispersionless limit $\hbar \to 0$. It satisfies 
an infinite number of highly nonlinear differential equations which follow
from the bilinear equations for the tau-function. The tau-function itself
does not make sense at $\hbar =0$.
 
For the purpose of performing the dispersionless limit 
the 4- and 2-point 
relations of the Hirota-Miwa type obtained in this paper seem to be 
necessary because they are especially suitable for this 
limit, as our experience on the examples of
the KP, DKP and Toda hierarchies teaches us.
(It is still unclear whether the dispersionless limit
can be performed just from integral bilinear relations of the form (\ref{main}),
without passing to relations of the Hirota-Miwa type.)

The dispersionless limit becomes especially interesting for
hierarchies of the Pfaff type. As it was first pointed out by Takasaki
in \cite{Takasaki07} on the example of the DKP (or coupled KP)
hierarchy, in its 
dispersionless version an elliptic curve naturally built in the 
structure of the hierarchy emerges. The elliptic modulus of this curve
is one of dynamical variables and depends on the hierarchical times.
Later, in \cite{AZ14}, this observation was taken 
as a starting point for elliptic 
parametrization (in terms of Jacobi theta-functions) of the dispersionless
hierarchy, in which its structure is significantly simplified 
(for the price of using elliptic functions instead of rational ones).
As was shown in \cite{Z24}, the dispersionless version of the 
multi-component KP hierarchy is related, in a similar way, to a
rational curve which is uniformized by trigonometric functions. 
Recently, we have shown \cite{SZ24} that the dispersionless 
multi-component DKP hierarchy admits an elliptic parametrization, too,
and this reformulation allows one to reduce the number of necessary 
equations considerably, making the structure of the hierarchy especially
transparent. The corresponding elliptic
curve is basically the same as the one that emerges in the one-component
DKP hierarchy, but in the multi-component case it enters the game
being equipped with some number of marked points, which are dynamical
variables, as well as the elliptic modulus. A generalization
of this picture to the multi-component Pfaff-Toda hierarchy in the
dispersionless limit will be addressed in our next paper. In this case
the expected reduction of the number of necessary equations is especially
important because, as we have seen, in the original
formulation there are too many of them to deal with 
(see the appendix for the full list
of 2-point equations).

As is known, the bilinear formalism based on the generating bilinear
functional relation for the tau-function is only one of possible 
approaches to the problem. Another fundamental approach is the 
Lax-Sato formalism based on dynamical equations for Lax operators
and representing the hierarchy in the form of equations of the
Zakharov-Shabat type. In this paper we have developed only the former,
leaving the latter for future research. A final goal in this respect
would be to create for the multi-component Pfaff-Toda hierarchy a 
general theory similar to the one developed in \cite{TZ25} for the
usual (not Pfaff) multi-component 2D Toda hierarchy. That means
treating the
system within both the bilinear and Lax-Sato formalisms, with a
proof of their equivalence.

\section*{Appendix: Two-point relations from the 4-point ones}

\addcontentsline{toc}{section}{Appendix: 
Two-point relations from the 4-point ones}
\def\theequation{A\arabic{equation}}
\def\theHequation{\theequation}
\setcounter{equation}{0}

\subsection*{Two-point relations that follow from (\ref{3010-compact})}

Let us demonstrate some details of 
the degeneration procedure on the example of
equation (\ref{3010-compact}) corresponding to the choice
$(L^ +, L^- | R^ +, R^- )=(3,0|1,0)$. There are two essentially
different options: $c,d\to \infty$ or $b,c\to \infty$. Below we 
consider them separately.

\paragraph{First option: $c,d \to \infty$.} There are different 
choices for the indices. The simplest one is $\alpha =\beta =\nu =\mu$,
which (after a shift of the variables) gives equation (\ref{eq1-1}).
In our short-hand notation it reads:
\beq\label{eq1-1a}
\begin{array}{l}
a \tau \Bigl ({\bf n}^{\alpha}, \bar {\bf n}^{\alpha}, {\bf t}^{[b_{\alpha}]},
\bar {\bf t}\Bigr )
\tau \Bigl ({\bf n}, \bar {\bf n}, {\bf t}^{[a_{\alpha}]},
\bar {\bf t}\Bigr )
-b \tau \Bigl ({\bf n}^{\alpha}, \bar {\bf n}^{\alpha}, {\bf t}^{[a_{\alpha}]},
\bar {\bf t}\Bigr )
\tau \Bigl ({\bf n}, \bar {\bf n}, {\bf t}^{[b_{\alpha}]},
\bar {\bf t}\Bigr )
\\ \\
=(a-b)\tau \Bigl ({\bf n}^{\alpha}, \bar {\bf n}^{\alpha}, 
{\bf t}^{[a_{\alpha}b_{\alpha}]},
\bar {\bf t}\Bigr )
\tau \Bigl ({\bf n}, \bar {\bf n}, 
{\bf t}, \bar {\bf t}\Bigr )
\\ \\
\phantom{aaaaaaaaaaaaaaaaaaaaa}
+(a^{-1}-b^{-1})
\tau \Bigl ({\bf n}^{\alpha \alpha}, \bar {\bf n}, 
{\bf t}^{[a_{\alpha}b_{\alpha}]},
\bar {\bf t}\Bigr )
\tau \Bigl ({\bf n}_{\alpha}, \bar {\bf n}^{\alpha}, {\bf t},
\bar {\bf t}\Bigr ).
\end{array}
\eeq
For the single $\alpha$th component, this equation is the same as 
in the one-component Pfaff-Toda hierarchy. 

Another choice is
$\alpha =\beta =\nu$ but $\mu \neq \alpha$.
Renaming $\mu \to \beta$ and shifting the arguments, we get the equation
\beq\label{eq2odd-1a}
\begin{array}{l}
a \tau \Bigl ({\bf n}^{\alpha}, \bar {\bf n}^{\beta}, {\bf t}^{[b_{\alpha}]},
\bar {\bf t}\Bigr )
\tau \Bigl ({\bf n}, \bar {\bf n}, {\bf t}^{[a_{\alpha}]},
\bar {\bf t}\Bigr )
-b \tau \Bigl ({\bf n}^{\alpha}, \bar {\bf n}^{\beta}, {\bf t}^{[a_{\alpha}]},
\bar {\bf t}\Bigr )
\tau \Bigl ({\bf n}, \bar {\bf n}, {\bf t}^{[b_{\alpha}]},
\bar {\bf t}\Bigr )
\\ \\
=(a-b)\tau \Bigl ({\bf n}^{\alpha}, \bar {\bf n}^{\beta}, 
{\bf t}^{[a_{\alpha}b_{\alpha}]},
\bar {\bf t}\Bigr )
\tau \Bigl ({\bf n}, \bar {\bf n}, {\bf t},
\bar {\bf t}\Bigr )
\\ \\
\phantom{aaaaaaaaaaaaaaaaaaaaaaa}
+(a^{-1}-b^{-1})
\tau \Bigl ({\bf n}^{\alpha \alpha}, \bar {\bf n}, 
{\bf t}^{[a_{\alpha}b_{\alpha}]},
\bar {\bf t}\Bigr )
\tau \Bigl ({\bf n}_{\alpha}, \bar {\bf n}^{\beta}, {\bf t},
\bar {\bf t}\Bigr )
\end{array}
\eeq
(at $\beta =\alpha$ it coincides with the previous one).

The next choice is
$\alpha =\beta =\mu$ but $\nu \neq \alpha$.
Renaming $\nu \to \beta$ and shifting the arguments, 
we get the equation
\beq\label{eq2odd-5a}
\begin{array}{l}
\tau \Bigl ({\bf n}^{\beta}, \bar {\bf n}^{\alpha}, {\bf t}^{[a_{\alpha}]},
\bar {\bf t}\Bigr )
\tau \Bigl ({\bf n}, \bar {\bf n}, {\bf t}^{[b_{\alpha}]},
\bar {\bf t}\Bigr )
-\tau \Bigl ({\bf n}^{\beta}, \bar {\bf n}^{\alpha}, {\bf t}^{[b_{\alpha}]},
\bar {\bf t}\Bigr )
\tau \Bigl ({\bf n}, \bar {\bf n}, {\bf t}^{[a_{\alpha}]},
\bar {\bf t}\Bigr )
\\ \\
\displaystyle{
=\epsilon_{\alpha \beta}(a^{-1}-b^{-1})\left [ \vphantom{\frac{a}{b}}
\tau \Bigl ({\bf n}^{\alpha \beta}, \bar {\bf n}, 
{\bf t}^{[a_{\alpha}b_{\alpha}]},
\bar {\bf t}\Bigr )
\tau \Bigl ({\bf n}_{\alpha}, \bar {\bf n}^{\alpha}, {\bf t},
\bar {\bf t}\Bigr ) \right. }
\\ \\
\displaystyle{\phantom{aaaaaaaaaaaaaaaaaaaaaa}
-\left. \vphantom{\frac{a}{b}}
\tau \Bigl ({\bf n}^{\alpha}, \bar {\bf n}^{\alpha}, 
{\bf t}^{[a_{\alpha}b_{\alpha}]},
\bar {\bf t}\Bigr )
\tau \Bigl ({\bf n}_{\alpha}^{\beta}, \bar {\bf n}, {\bf t},
\bar {\bf t}\Bigr ) \right ]. }
\end{array}
\eeq

The next possible choice is
$\alpha =\beta$, $\nu =\mu$ but $\nu \neq \alpha$.
Renaming $\nu \to \beta$ and shifting the arguments, 
we get the equation
\beq\label{eq2odd-6a}
\begin{array}{l}
\tau \Bigl ({\bf n}^{\beta}, \bar {\bf n}^{\beta}, {\bf t}^{[a_{\alpha}]},
\bar {\bf t}\Bigr )
\tau \Bigl ({\bf n}, \bar {\bf n}, {\bf t}^{[b_{\alpha}]},
\bar {\bf t}\Bigr )
-\tau \Bigl ({\bf n}^{\beta}, \bar {\bf n}^{\beta}, {\bf t}^{[b_{\alpha}]},
\bar {\bf t}\Bigr )
\tau \Bigl ({\bf n}, \bar {\bf n}, {\bf t}^{[a_{\alpha}]},
\bar {\bf t}\Bigr )
\\ \\
\displaystyle{
=\epsilon_{\alpha \beta}(a^{-1}-b^{-1})\left [ \vphantom{\frac{a}{b}}
\tau \Bigl ({\bf n}^{\alpha \beta}, \bar {\bf n}, 
{\bf t}^{[a_{\alpha}b_{\alpha}]},
\bar {\bf t}\Bigr )
\tau \Bigl ({\bf n}_{\alpha}, \bar {\bf n}^{\beta}, {\bf t},
\bar {\bf t}\Bigr ) \right. }
\\ \\
\displaystyle{\phantom{aaaaaaaaaaaaaaaaaaaaaa}
-\left. \vphantom{\frac{a}{b}}
\tau \Bigl ({\bf n}^{\alpha}, \bar {\bf n}^{\beta}, 
{\bf t}^{[a_{\alpha}b_{\alpha}]},
\bar {\bf t}\Bigr )
\tau \Bigl ({\bf n}_{\alpha}^{\beta}, \bar {\bf n}, {\bf t},
\bar {\bf t}\Bigr ) \right ]. }
\end{array}
\eeq

Another choice is $\alpha =\nu =\mu$, but $\beta\neq \alpha$,
which leads to the equation
\beq\label{eq3odd-1a}
\begin{array}{l}
\tau \Bigl ({\bf n}^{\beta}, \bar {\bf n}^{\alpha}, {\bf t}^{[b_{\beta}]},
\bar {\bf t}\Bigr )
\tau \Bigl ({\bf n}, \bar {\bf n}, {\bf t}^{[a_{\alpha}]},
\bar {\bf t}\Bigr )
-\tau \Bigl ({\bf n}^{\beta}, \bar {\bf n}^{\alpha}, 
{\bf t}^{[a_{\alpha}b_{\beta}]},
\bar {\bf t}\Bigr )
\tau \Bigl ({\bf n}, \bar {\bf n}, {\bf t},
\bar {\bf t}\Bigr )
\\ \\
\displaystyle{
=\epsilon_{\alpha \beta}a^{-1}\left [ \vphantom{\frac{a}{b}}
\tau \Bigl ({\bf n}^{\alpha}, \bar {\bf n}^{\alpha}, 
{\bf t}^{[a_{\alpha}]},
\bar {\bf t}\Bigr )
\tau \Bigl ({\bf n}_{\alpha}^{\beta}, \bar {\bf n}, {\bf t}^{[b_{\beta}]},
\bar {\bf t}\Bigr ) -
\tau \Bigl ({\bf n}^{\alpha \beta}, \bar {\bf n}, 
{\bf t}^{[a_{\alpha}b_{\beta}]},
\bar {\bf t}\Bigr )
\tau \Bigl ({\bf n}_{\alpha}, \bar {\bf n}^{\alpha}, {\bf t},
\bar {\bf t}\Bigr ) \right ]. }
\end{array}
\eeq

The last possible choice is $\alpha =\nu$, $\beta = \mu$,
keeping $\beta\neq \alpha$. The corresponding equation reads
\beq\label{eq3odd-2a}
\begin{array}{l}
\tau \Bigl ({\bf n}^{\beta}, \bar {\bf n}^{\beta}, {\bf t}^{[b_{\beta}]},
\bar {\bf t}\Bigr )
\tau \Bigl ({\bf n}, \bar {\bf n}, {\bf t}^{[a_{\alpha}]},
\bar {\bf t}\Bigr )
-\tau \Bigl ({\bf n}^{\beta}, \bar {\bf n}^{\beta}, 
{\bf t}^{[a_{\alpha}b_{\beta}]},
\bar {\bf t}\Bigr )
\tau \Bigl ({\bf n}, \bar {\bf n}, {\bf t},
\bar {\bf t}\Bigr )
\\ \\
\displaystyle{
=\epsilon_{\alpha \beta}a^{-1}\left [ \vphantom{\frac{a}{b}}
\tau \Bigl ({\bf n}^{\alpha}, \bar {\bf n}^{\beta}, 
{\bf t}^{[a_{\alpha}]},
\bar {\bf t}\Bigr )
\tau \Bigl ({\bf n}_{\alpha}^{\beta}, \bar {\bf n}, {\bf t}^{[b_{\beta}]},
\bar {\bf t}\Bigr ) -
\tau \Bigl ({\bf n}^{\alpha \beta}, \bar {\bf n}, 
{\bf t}^{[a_{\alpha}b_{\beta}]},
\bar {\bf t}\Bigr )
\tau \Bigl ({\bf n}_{\alpha}, \bar {\bf n}^{\beta}, {\bf t},
\bar {\bf t}\Bigr ) \right ]. }
\end{array}
\eeq
In fact there are some other choices but they are equivalent to the ones
considered above.

\paragraph{Second option: $b=c \to \infty$.} In this case, if
in addition $\beta =\nu$,
equation (\ref{3010-compact}) converts into trivial identity.
Hence we first assume that $\beta \neq \nu$ and tend $b,c \to \infty$
afterwards. 

The first choice is
$\alpha =\nu =\mu$ with $\alpha \neq \beta$.
Renaming some indices and points,
and shifting the arguments, 
we get the equation
\beq\label{eq4odd-1a}
\begin{array}{l}
\tau \Bigl ({\bf n}^{\beta}, \bar {\bf n}^{\alpha}, {\bf t},
\bar {\bf t}^{[b_{\alpha}]}\Bigr )
\tau \Bigl ({\bf n}, \bar {\bf n}, {\bf t}^{[a_{\alpha}]},
\bar {\bf t}\Bigr )
-\tau \Bigl ({\bf n}^{\beta}, \bar {\bf n}^{\alpha}, 
{\bf t}^{[a_{\alpha}]},
\bar {\bf t}^{[b_{\alpha}]} \Bigr )
\tau \Bigl ({\bf n}, \bar {\bf n}, {\bf t},
\bar {\bf t}\Bigr )
\\ \\
\displaystyle{
=\epsilon_{\alpha \beta}a^{-1}\left [ \vphantom{\frac{a}{b}}
\tau \Bigl ({\bf n}^{\alpha}, \bar {\bf n}^{\alpha}, 
{\bf t}^{[a_{\alpha}]},
\bar {\bf t}^{[b_{\alpha}]}\Bigr )
\tau \Bigl ({\bf n}_{\alpha}^{\beta}, \bar {\bf n}, {\bf t},
\bar {\bf t}\Bigr ) -
\tau \Bigl ({\bf n}^{\alpha \beta}, \bar {\bf n}, 
{\bf t}^{[a_{\alpha}]},
\bar {\bf t}\Bigr )
\tau \Bigl ({\bf n}_{\alpha}, \bar {\bf n}^{\alpha}, {\bf t},
\bar {\bf t}^{[b_{\alpha}]}\Bigr ) \right ]. }
\end{array}
\eeq
Next, the choice
$\alpha =\nu $, $\beta =\mu$ with $\alpha \neq \beta$ leads to
the equation
\beq\label{eq5odd-1a}
\begin{array}{l}
\tau \Bigl ({\bf n}^{\beta}, \bar {\bf n}^{\beta}, {\bf t},
\bar {\bf t}^{[b_{\beta}]}\Bigr )
\tau \Bigl ({\bf n}, \bar {\bf n}, {\bf t}^{[a_{\alpha}]},
\bar {\bf t}\Bigr )
-\tau \Bigl ({\bf n}^{\beta}, \bar {\bf n}^{\beta}, 
{\bf t}^{[a_{\alpha}]},
\bar {\bf t}^{[b_{\beta}]} \Bigr )
\tau \Bigl ({\bf n}, \bar {\bf n}, {\bf t},
\bar {\bf t}\Bigr )
\\ \\
\displaystyle{
=\epsilon_{\alpha \beta}a^{-1}\left [ \vphantom{\frac{a}{b}}
\tau \Bigl ({\bf n}^{\alpha}, \bar {\bf n}^{\beta}, 
{\bf t}^{[a_{\alpha}]},
\bar {\bf t}^{[b_{\beta}]}\Bigr )
\tau \Bigl ({\bf n}_{\alpha}^{\beta}, \bar {\bf n}, {\bf t},
\bar {\bf t}\Bigr ) -
\tau \Bigl ({\bf n}^{\alpha \beta}, \bar {\bf n}, 
{\bf t}^{[a_{\alpha}]},
\bar {\bf t}\Bigr )
\tau \Bigl ({\bf n}_{\alpha}, \bar {\bf n}^{\beta}, {\bf t},
\bar {\bf t}^{[b_{\beta}]}\Bigr ) \right ]. }
\end{array}
\eeq
At last, the choice
$\alpha =\mu$, $\beta =\nu$ with $\alpha \neq \beta$ leads to
the same equation (\ref{eq5odd-1a}).

Since the points $a,b,c$, together with the corresponding indices,
enter the game in a symmetric manner, other options (like $b,d\to \infty$, 
$a,b \to \infty$, etc.)  are equivalent to one of 
the two considered above.

We have seen that degenerations of the 4-point relation corresponding to the
$$(L^ +, L^- | R^ +, R^- )=(3,0|1,0)$$ scheme yields 8 non-equivalent 
2-point relations\footnote{Some of them can be obtained from each other
by changing signs of some arguments and shifts: $\bar {\bf n}\to -\bar {\bf n}$,
etc.}.

\subsection*{Completing the list of two-point relations}

It turns out that each of the three remaining 4-point relations give, after
different degenerations,
$8$ two-point ones (obeying condition C), so total number of them is 32.
The analysis is very similar to the one given above for the scheme
$(3,0|1,0)$, so we omit the details and present here only the remaining
items of the full list
of the two-point relations. 

\subsubsection*{Two-point relations that follow from 
$(L^ +, L^- | R^ +, R^- )=(3,0|0,1)$}

\beq\label{eq1-2a}
\begin{array}{ll}
\bullet & a \tau \Bigl ({\bf n}^{\alpha}, \bar {\bf n}, {\bf t}^{[b_{\alpha}]},
\bar {\bf t}\Bigr )
\tau \Bigl ({\bf n}, \bar {\bf n}^{\alpha}, {\bf t}^{[a_{\alpha}]},
\bar {\bf t}\Bigr )
-b \tau \Bigl ({\bf n}^{\alpha}, \bar {\bf n}, {\bf t}^{[a_{\alpha}]},
\bar {\bf t}\Bigr )
\tau \Bigl ({\bf n}, \bar {\bf n}^{\alpha}, {\bf t}^{[b_{\alpha}]},
\bar {\bf t}\Bigr )
\\ &\\
&=(a-b)\tau \Bigl ({\bf n}^{\alpha}, \bar {\bf n}, 
{\bf t}^{[a_{\alpha}b_{\alpha}]},
\bar {\bf t}\Bigr )
\tau \Bigl ({\bf n}, \bar {\bf n}^{\alpha}, 
{\bf t}, \bar {\bf t}\Bigr )
\\ &\\
&\phantom{aaaaaaaaaaaaaaaaaaa}
+(a^{-1}-b^{-1})
\tau \Bigl ({\bf n}^{\alpha \alpha}, \bar {\bf n}^{\alpha}, 
{\bf t}^{[a_{\alpha}b_{\alpha}]},
\bar {\bf t}\Bigr )
\tau \Bigl ({\bf n}_{\alpha}, \bar {\bf n}, {\bf t},
\bar {\bf t}\Bigr ),
\end{array}
\eeq

\beq\label{eq2odd-2a}
\begin{array}{ll}
\bullet & a \tau \Bigl ({\bf n}^{\alpha}, \bar {\bf n}, {\bf t}^{[b_{\alpha}]},
\bar {\bf t}\Bigr )
\tau \Bigl ({\bf n}, \bar {\bf n}^{\beta}, {\bf t}^{[a_{\alpha}]},
\bar {\bf t}\Bigr )
-b \tau \Bigl ({\bf n}^{\alpha}, \bar {\bf n}, {\bf t}^{[a_{\alpha}]},
\bar {\bf t}\Bigr )
\tau \Bigl ({\bf n}, \bar {\bf n}^{\beta}, {\bf t}^{[b_{\alpha}]},
\bar {\bf t}\Bigr )
\\ &\\
&=(a-b)\tau \Bigl ({\bf n}^{\alpha}, \bar {\bf n}, 
{\bf t}^{[a_{\alpha}b_{\alpha}]},
\bar {\bf t}\Bigr )
\tau \Bigl ({\bf n}, \bar {\bf n}^{\beta}, 
{\bf t}, \bar {\bf t}\Bigr )
\\ \\
&\phantom{aaaaaaaaaaaaaaaaaaa}
+(a^{-1}-b^{-1})
\tau \Bigl ({\bf n}^{\alpha \alpha}, \bar {\bf n}^{\beta}, 
{\bf t}^{[a_{\alpha}b_{\alpha}]},
\bar {\bf t}\Bigr )
\tau \Bigl ({\bf n}_{\alpha}, \bar {\bf n}, {\bf t},
\bar {\bf t}\Bigr ),
\end{array}
\eeq

\beq\label{eq2odd-8a}
\begin{array}{ll}
\bullet &
\tau \Bigl ({\bf n}^{\beta}, \bar {\bf n}, {\bf t}^{[a_{\alpha}]},
\bar {\bf t}\Bigr )
\tau \Bigl ({\bf n}, \bar {\bf n}^{\alpha}, {\bf t}^{[b_{\alpha}]},
\bar {\bf t}\Bigr )
-\tau \Bigl ({\bf n}^{\beta}, \bar {\bf n}, {\bf t}^{[b_{\alpha}]},
\bar {\bf t}\Bigr )
\tau \Bigl ({\bf n}, \bar {\bf n}^{\alpha}, {\bf t}^{[a_{\alpha}]},
\bar {\bf t}\Bigr )
\\ &\\
&\displaystyle{
=\epsilon_{\alpha \beta}(a^{-1}-b^{-1})\left [ \vphantom{\frac{a}{b}}
\tau \Bigl ({\bf n}^{\alpha \beta}, \bar {\bf n}^{\alpha}, 
{\bf t}^{[a_{\alpha}b_{\alpha}]},
\bar {\bf t}\Bigr )
\tau \Bigl ({\bf n}_{\alpha}, \bar {\bf n}, {\bf t},
\bar {\bf t}\Bigr ) \right. }
\\ &\\
&\displaystyle{\phantom{aaaaaaaaaaaaaaaaaaaaaa}
-\left. \vphantom{\frac{a}{b}}
\tau \Bigl ({\bf n}^{\alpha}, \bar {\bf n}, 
{\bf t}^{[a_{\alpha}b_{\alpha}]},
\bar {\bf t}\Bigr )
\tau \Bigl ({\bf n}_{\alpha}^{\beta}, \bar {\bf n}^{\alpha}, {\bf t},
\bar {\bf t}\Bigr ) \right ], }
\end{array}
\eeq

\beq\label{eq2odd-7a}
\begin{array}{ll}
\bullet &
\tau \Bigl ({\bf n}^{\beta}, \bar {\bf n}, {\bf t}^{[a_{\alpha}]},
\bar {\bf t}\Bigr )
\tau \Bigl ({\bf n}, \bar {\bf n}^{\beta}, {\bf t}^{[b_{\alpha}]},
\bar {\bf t}\Bigr )
-\tau \Bigl ({\bf n}^{\beta}, \bar {\bf n}, {\bf t}^{[b_{\alpha}]},
\bar {\bf t}\Bigr )
\tau \Bigl ({\bf n}, \bar {\bf n}^{\beta}, {\bf t}^{[a_{\alpha}]},
\bar {\bf t}\Bigr )
\\ &\\
&\displaystyle{
=\epsilon_{\alpha \beta}(a^{-1}-b^{-1})\left [ \vphantom{\frac{a}{b}}
\tau \Bigl ({\bf n}^{\alpha \beta}, \bar {\bf n}^{\beta}, 
{\bf t}^{[a_{\alpha}b_{\alpha}]},
\bar {\bf t}\Bigr )
\tau \Bigl ({\bf n}_{\alpha}, \bar {\bf n}, {\bf t},
\bar {\bf t}\Bigr ) \right. }
\\ &\\
&\displaystyle{\phantom{aaaaaaaaaaaaaaaaaaaaaa}
-\left. \vphantom{\frac{a}{b}}
\tau \Bigl ({\bf n}^{\alpha}, \bar {\bf n}, 
{\bf t}^{[a_{\alpha}b_{\alpha}]},
\bar {\bf t}\Bigr )
\tau \Bigl ({\bf n}_{\alpha}^{\beta}, \bar {\bf n}^{\beta}, {\bf t},
\bar {\bf t}\Bigr ) \right ], }
\end{array}
\eeq

\beq\label{eq3odd-4a}
\begin{array}{ll}
\bullet &
\tau \Bigl ({\bf n}^{\beta}, \bar {\bf n}, {\bf t}^{[b_{\beta}]},
\bar {\bf t}\Bigr )
\tau \Bigl ({\bf n}, \bar {\bf n}^{\alpha}, {\bf t}^{[a_{\alpha}]},
\bar {\bf t}\Bigr )
-\tau \Bigl ({\bf n}^{\beta}, \bar {\bf n}, {\bf t}^{[a_{\alpha}b_{\beta}]},
\bar {\bf t}\Bigr )
\tau \Bigl ({\bf n}, \bar {\bf n}^{\alpha}, {\bf t},
\bar {\bf t}\Bigr )
\\ &\\
&\displaystyle{
=\epsilon_{\alpha \beta}a^{-1}\left [ \vphantom{\frac{a}{b}}
\tau \Bigl ({\bf n}^{\alpha }, \bar {\bf n}, 
{\bf t}^{[a_{\alpha}]},
\bar {\bf t}\Bigr )
\tau \Bigl ({\bf n}_{\alpha}^{\beta}, \bar {\bf n}^{\alpha}, {\bf t}^{[b_{\beta}]},
\bar {\bf t}\Bigr ) -
\tau \Bigl ({\bf n}^{\alpha \beta}, \bar {\bf n}^{\alpha}, 
{\bf t}^{[a_{\alpha}b_{\beta}]},
\bar {\bf t}\Bigr )
\tau \Bigl ({\bf n}_{\alpha}, \bar {\bf n}, {\bf t},
\bar {\bf t}\Bigr ) \right ], }
\end{array}
\eeq

\beq\label{eq3odd-3a}
\begin{array}{ll}
\bullet &
\tau \Bigl ({\bf n}^{\beta}, \bar {\bf n}, {\bf t}^{[b_{\beta}]},
\bar {\bf t}\Bigr )
\tau \Bigl ({\bf n}, \bar {\bf n}^{\beta}, {\bf t}^{[a_{\alpha}]},
\bar {\bf t}\Bigr )
-\tau \Bigl ({\bf n}^{\beta}, \bar {\bf n}, {\bf t}^{[a_{\alpha}b_{\beta}]},
\bar {\bf t}\Bigr )
\tau \Bigl ({\bf n}, \bar {\bf n}^{\beta}, {\bf t},
\bar {\bf t}\Bigr )
\\ &\\
&\displaystyle{
=\epsilon_{\alpha \beta}a^{-1}\left [ \vphantom{\frac{a}{b}}
\tau \Bigl ({\bf n}^{\alpha }, \bar {\bf n}, 
{\bf t}^{[a_{\alpha}]},
\bar {\bf t}\Bigr )
\tau \Bigl ({\bf n}_{\alpha}^{\beta}, \bar {\bf n}^{\beta}, {\bf t}^{[b_{\beta}]},
\bar {\bf t}\Bigr ) 
-\tau \Bigl ({\bf n}^{\alpha \beta}, \bar {\bf n}^{\beta}, 
{\bf t}^{[a_{\alpha}b_{\beta}]},
\bar {\bf t}\Bigr )
\tau \Bigl ({\bf n}_{\alpha}, \bar {\bf n}, {\bf t},
\bar {\bf t}\Bigr ) \right ], }
\end{array}
\eeq

\beq\label{eq4odd-2a}
\begin{array}{ll}
\bullet &
\tau \Bigl ({\bf n}^{\beta}, \bar {\bf n}, {\bf t}^{[a_{\alpha}]},
\bar {\bf t}\Bigr )
\tau \Bigl ({\bf n}, \bar {\bf n}^{\alpha}, {\bf t},
\bar {\bf t}^{[b_{\alpha}]}\Bigr )
-\tau \Bigl ({\bf n}^{\beta}, \bar {\bf n}, {\bf t},
\bar {\bf t}\Bigr )
\tau \Bigl ({\bf n}, \bar {\bf n}^{\alpha}, {\bf t}^{[a_{\alpha}]},
\bar {\bf t}^{[b_{\alpha}]}\Bigr )
\\ &\\
&\displaystyle{
=\epsilon_{\alpha \beta}a^{-1}\left [ \vphantom{\frac{a}{b}}
\tau \Bigl ({\bf n}^{\alpha \beta}, \bar {\bf n}^{\alpha}, 
{\bf t}^{[a_{\alpha}]},
\bar {\bf t}^{[b_{\alpha}]}\Bigr )
\tau \Bigl ({\bf n}_{\alpha}, \bar {\bf n}, {\bf t},
\bar {\bf t}\Bigr ) -
\tau \Bigl ({\bf n}^{\alpha}, \bar {\bf n}, 
{\bf t}^{[a_{\alpha}]},
\bar {\bf t}\Bigr )
\tau \Bigl ({\bf n}_{\alpha}^{\beta}, \bar {\bf n}^{\alpha}, {\bf t},
\bar {\bf t}^{[b_{\alpha}]}\Bigr ) \right ], }
\end{array}
\eeq

\beq\label{eq5odd-2a}
\begin{array}{ll}
\bullet &
\tau \Bigl ({\bf n}^{\beta}, \bar {\bf n}, {\bf t}^{[a_{\alpha}]},
\bar {\bf t}\Bigr )
\tau \Bigl ({\bf n}, \bar {\bf n}^{\beta}, {\bf t},
\bar {\bf t}^{[b_{\beta}]}\Bigr )
-\tau \Bigl ({\bf n}^{\beta}, \bar {\bf n}, {\bf t},
\bar {\bf t}\Bigr )
\tau \Bigl ({\bf n}, \bar {\bf n}^{\beta}, {\bf t}^{[a_{\alpha}]},
\bar {\bf t}^{[b_{\alpha}]}\Bigr )
\\ &\\
&
=\epsilon_{\alpha \beta} \left [ \vphantom{\frac{a}{b}}
\tau \Bigl ({\bf n}^{\alpha \beta}, \bar {\bf n}^{\beta}, 
{\bf t}^{[a_{\alpha}]},
\bar {\bf t}^{[b_{\alpha}]}\Bigr )
\tau \Bigl ({\bf n}_{\alpha}, \bar {\bf n}, {\bf t},
\bar {\bf t}\Bigr ) -
\tau \Bigl ({\bf n}^{\alpha}, \bar {\bf n}, 
{\bf t}^{[a_{\alpha}]},
\bar {\bf t}\Bigr )
\tau \Bigl ({\bf n}_{\alpha}^{\beta}, \bar {\bf n}^{\beta}, {\bf t},
\bar {\bf t}^{[b_{\beta}]}\Bigr )\right ].
\end{array}
\eeq

\subsubsection*{Two-point relations that follow from 
$(L^ +, L^- | R^ +, R^- )=(2,0|2,0)$}

\beq\label{eq3-01a}
\begin{array}{ll}
\bullet & a \tau \Bigl ({\bf n}, \bar {\bf n}^{\alpha}, {\bf t},
\bar {\bf t}^{[b_{\alpha}]}\Bigr )
\tau \Bigl ({\bf n}, \bar {\bf n}_{\alpha}, {\bf t}^{[a_{\alpha}]},
\bar {\bf t}\Bigr )
-a \tau \Bigl ({\bf n}, \bar {\bf n}^{\alpha}, {\bf t}^{[a_{\alpha}]},
\bar {\bf t}^{[b_{\alpha}]}\Bigr )
\tau \Bigl ({\bf n}, \bar {\bf n}_{\alpha}, {\bf t},
\bar {\bf t}\Bigr )
\\ &\\
&=b\tau \Bigl ({\bf n}^{\alpha}, \bar {\bf n}, 
{\bf t}^{[a_{\alpha}]},
\bar {\bf t}\Bigr )
\tau \Bigl ({\bf n}_{\alpha}, \bar {\bf n}, 
{\bf t}, \bar {\bf t}^{[b_{\alpha}]}\Bigr )
-b
\tau \Bigl ({\bf n}^{\alpha}, \bar {\bf n}, 
{\bf t}^{[a_{\alpha}]},
\bar {\bf t}^{[b_{\alpha}]}\Bigr )
\tau \Bigl ({\bf n}_{\alpha}, \bar {\bf n}, {\bf t},
\bar {\bf t}\Bigr ),
\end{array}
\eeq

\beq\label{eq4even-1a}
\begin{array}{ll}
\bullet &
\tau \Bigl ({\bf n}, \bar {\bf n}^{\alpha \beta}, {\bf t},
\bar {\bf t}^{[b_{\alpha}]}\Bigr )
\tau \Bigl ({\bf n}, \bar {\bf n}, {\bf t}^{[a_{\alpha}]},
\bar {\bf t}\Bigr )
-\tau \Bigl ({\bf n}, \bar {\bf n}^{\alpha \beta}, {\bf t}^{[a_{\alpha}]},
\bar {\bf t}^{[b_{\alpha}]}\Bigr )
\tau \Bigl ({\bf n}, \bar {\bf n}, {\bf t},
\bar {\bf t}\Bigr )
\\ &\\
&\displaystyle{
=\epsilon_{\alpha \beta}a^{-1}\left [ \vphantom{\frac{a}{b}}
\tau \Bigl ({\bf n}^{\alpha}, \bar {\bf n}^{\alpha}, 
{\bf t}^{[a_{\alpha}]},
\bar {\bf t}^{[b_{\alpha}]}\Bigr )
\tau \Bigl ({\bf n}_{\alpha}, \bar {\bf n}^{\beta}, {\bf t},
\bar {\bf t}\Bigr ) -
\tau \Bigl ({\bf n}^{\alpha}, \bar {\bf n}^{\beta}, 
{\bf t}^{[a_{\alpha}]},
\bar {\bf t}\Bigr )
\tau \Bigl ({\bf n}_{\alpha}, \bar {\bf n}^{\alpha}, {\bf t},
\bar {\bf t}^{[b_{\alpha}]}\Bigr ) \right ], }
\end{array}
\eeq

\beq\label{eq4even-3a}
\begin{array}{ll}
\bullet &
\tau \Bigl ({\bf n}^{\alpha}, \bar {\bf n}^{\alpha \beta}, {\bf t}^{[a_{\alpha}]},
\bar {\bf t}^{[b_{\beta}]}\Bigr )
\tau \Bigl ({\bf n}^{\beta}, \bar {\bf n}, {\bf t},
\bar {\bf t}\Bigr )
-\tau \Bigl ({\bf n}^{\alpha \beta}, \bar {\bf n}^{\alpha}, {\bf t}^{[a_{\alpha}]},
\bar {\bf t}^{[b_{\alpha}]}\Bigr )
\tau \Bigl ({\bf n}, \bar {\bf n}^{\beta}, {\bf t},
\bar {\bf t}\Bigr )
\\ &\\
&
=\tau \Bigl ({\bf n}^{\beta}, \bar {\bf n}^{\alpha \beta}, 
{\bf t},
\bar {\bf t}^{[b_{\alpha}]}\Bigr )
\tau \Bigl ({\bf n}^{\alpha}, \bar {\bf n}, {\bf t}^{[a_{\alpha}]},
\bar {\bf t}\Bigr ) -
\tau \Bigl ({\bf n}^{\alpha \beta}, \bar {\bf n}^{\beta}, 
{\bf t}^{[a_{\alpha}]},
\bar {\bf t}\Bigr )
\tau \Bigl ({\bf n}, \bar {\bf n}^{\alpha}, {\bf t},
\bar {\bf t}^{[b_{\alpha}]}\Bigr ).
\end{array}
\eeq

\beq\label{eq5even-1a}
\begin{array}{ll}
\bullet &
\tau \Bigl ({\bf n}, \bar {\bf n}^{\alpha \beta}, {\bf t},
\bar {\bf t}^{[b_{\beta}]}\Bigr )
\tau \Bigl ({\bf n}, \bar {\bf n}, {\bf t}^{[a_{\alpha}]},
\bar {\bf t}\Bigr )
-\tau \Bigl ({\bf n}, \bar {\bf n}^{\alpha \beta}, {\bf t}^{[a_{\alpha}]},
\bar {\bf t}^{[b_{\beta}]}\Bigr )
\tau \Bigl ({\bf n}, \bar {\bf n}, {\bf t},
\bar {\bf t}\Bigr )
\\ &\\
&\displaystyle{
=\epsilon_{\alpha \beta}a^{-1}\left [ \vphantom{\frac{a}{b}}
\tau \Bigl ({\bf n}^{\alpha}, \bar {\bf n}^{\alpha}, 
{\bf t}^{[a_{\alpha}]},
\bar {\bf t}\Bigr )
\tau \Bigl ({\bf n}_{\alpha}, \bar {\bf n}^{\beta}, {\bf t},
\bar {\bf t}^{[b_{\beta}]}\Bigr ) -
\tau \Bigl ({\bf n}^{\alpha}, \bar {\bf n}^{\beta}, 
{\bf t}^{[a_{\alpha}]},
\bar {\bf t}^{[b_{\beta}]}\Bigr )
\tau \Bigl ({\bf n}_{\alpha}, \bar {\bf n}^{\alpha}, {\bf t},
\bar {\bf t}\Bigr ) \right ], }
\end{array}
\eeq

\beq\label{eq5even-2a}
\begin{array}{ll}
\bullet & a \tau \Bigl ({\bf n}, \bar {\bf n}^{\beta}, {\bf t},
\bar {\bf t}^{[b_{\beta}]}\Bigr )
\tau \Bigl ({\bf n}, \bar {\bf n}_{\beta}, {\bf t}^{[a_{\alpha}]},
\bar {\bf t}\Bigr )
-a \tau \Bigl ({\bf n}, \bar {\bf n}^{\beta}, {\bf t}^{[a_{\alpha}]},
\bar {\bf t}^{[b_{\alpha}]}\Bigr )
\tau \Bigl ({\bf n}, \bar {\bf n}_{\beta}, {\bf t},
\bar {\bf t}\Bigr )
\\ &\\
&=b\tau \Bigl ({\bf n}^{\alpha}, \bar {\bf n}, 
{\bf t}^{[a_{\alpha}]},
\bar {\bf t}\Bigr )
\tau \Bigl ({\bf n}_{\alpha}, \bar {\bf n}, 
{\bf t}, \bar {\bf t}^{[b_{\beta}]}\Bigr )
-b
\tau \Bigl ({\bf n}^{\alpha}, \bar {\bf n}, 
{\bf t}^{[a_{\alpha}]},
\bar {\bf t}^{[b_{\alpha}]}\Bigr )
\tau \Bigl ({\bf n}_{\alpha}, \bar {\bf n}, {\bf t},
\bar {\bf t}\Bigr ),
\end{array}
\eeq

\beq\label{eq5even-5a}
\begin{array}{ll}
\bullet &
\tau \Bigl ({\bf n}^{\alpha}, \bar {\bf n}^{\alpha \beta}, {\bf t}^{[a_{\alpha}]},
\bar {\bf t}^{[b_{\beta}]}\Bigr )
\tau \Bigl ({\bf n}^{\beta}, \bar {\bf n}, {\bf t},
\bar {\bf t}\Bigr )
-\tau \Bigl ({\bf n}^{\beta}, \bar {\bf n}^{\alpha \beta}, {\bf t},
\bar {\bf t}^{[b_{\alpha}]}\Bigr )
\tau \Bigl ({\bf n}^{\alpha}, \bar {\bf n}, {\bf t}^{[a_{\alpha}]},
\bar {\bf t}\Bigr )
\\ &\\
&
=\tau \Bigl ({\bf n}^{\alpha \beta}, \bar {\bf n}^{\alpha}, 
{\bf t}^{[a_{\alpha}]},
\bar {\bf t}\Bigr )
\tau \Bigl ({\bf n}, \bar {\bf n}^{\beta}, {\bf t},
\bar {\bf t}^{[b_{\beta}]}\Bigr ) -
\tau \Bigl ({\bf n}^{\alpha \beta}, \bar {\bf n}^{\beta}, 
{\bf t}^{[a_{\alpha}]},
\bar {\bf t}^{[b_{\beta}]}\Bigr )
\tau \Bigl ({\bf n}, \bar {\bf n}^{\alpha}, {\bf t},
\bar {\bf t}\Bigr ),
\end{array}
\eeq

\beq\label{eq2even-1a}
\begin{array}{ll}
\bullet &
\tau \Bigl ({\bf n}, \bar {\bf n}^{\alpha \beta}, {\bf t}^{[a_{\alpha}]},
\bar {\bf t}\Bigr )
\tau \Bigl ({\bf n}, \bar {\bf n}, {\bf t}^{[b_{\alpha}]},
\bar {\bf t}\Bigr )
-\tau \Bigl ({\bf n}, \bar {\bf n}^{\alpha \beta}, {\bf t}^{[b_{\alpha}]},
\bar {\bf t}\Bigr )
\tau \Bigl ({\bf n}, \bar {\bf n}, {\bf t}^{[a_{\alpha}]},
\bar {\bf t}\Bigr )
\\ &\\
&\displaystyle{
=\epsilon_{\alpha \beta}(a^{-1}-b^{-1})\left [ \vphantom{\frac{a}{b}}
\tau \Bigl ({\bf n}^{\alpha}, \bar {\bf n}^{\beta}, 
{\bf t}^{[a_{\alpha}b_{\alpha}]},
\bar {\bf t}\Bigr )
\tau \Bigl ({\bf n}_{\alpha}, \bar {\bf n}^{\alpha}, {\bf t},
\bar {\bf t}\Bigr ) \right. }
\\ &\\
&\displaystyle{\phantom{aaaaaaaaaaaaaaaaaaaaaa}
-\left. \vphantom{\frac{a}{b}}
\tau \Bigl ({\bf n}^{\alpha}, \bar {\bf n}^{\alpha}, 
{\bf t}^{[a_{\alpha}b_{\alpha}]},
\bar {\bf t}\Bigr )
\tau \Bigl ({\bf n}_{\alpha}, \bar {\bf n}^{\beta}, {\bf t},
\bar {\bf t}\Bigr ) \right ], }
\end{array}
\eeq

\beq\label{eq3even-1a}
\begin{array}{ll}
\bullet &
\tau \Bigl ({\bf n}^{\alpha}, \bar {\bf n}^{\alpha \beta}, {\bf t}^{[a_{\alpha}]},
\bar {\bf t}\Bigr )
\tau \Bigl ({\bf n}^{\beta}, \bar {\bf n}, {\bf t}^{[b_{\beta}]},
\bar {\bf t}\Bigr )
-\tau \Bigl ({\bf n}^{\beta}, \bar {\bf n}^{\alpha \beta}, {\bf t}^{[b_{\beta}]},
\bar {\bf t}\Bigr )
\tau \Bigl ({\bf n}^{\alpha}, \bar {\bf n}, {\bf t}^{[a_{\alpha}]},
\bar {\bf t}\Bigr )
\\ &\\
&
=\tau \Bigl ({\bf n}^{\alpha \beta}, \bar {\bf n}^{\alpha}, 
{\bf t}^{[a_{\alpha}b_{\beta}]},
\bar {\bf t}\Bigr )
\tau \Bigl ({\bf n}, \bar {\bf n}^{\beta}, {\bf t},
\bar {\bf t}\Bigr ) -
\tau \Bigl ({\bf n}^{\alpha \beta}, \bar {\bf n}^{\beta}, 
{\bf t}^{[a_{\alpha}b_{\beta}]},
\bar {\bf t}\Bigr )
\tau \Bigl ({\bf n}, \bar {\bf n}^{\alpha}, {\bf t},
\bar {\bf t}\Bigr ).
\end{array}
\eeq

\subsubsection*{Two-point relations that follow from 
$(L^ +, L^- | R^ +, R^- )=(2,0|1,1)$}

\beq\label{eq3-02a}
\begin{array}{ll}
\bullet &
\tau \Bigl ({\bf n}, \bar {\bf n}, {\bf t}^{[a_{\alpha}]},
\bar {\bf t}\Bigr )
\tau \Bigl ({\bf n}, \bar {\bf n}, {\bf t},
\bar {\bf t}^{[b_{\alpha}]}\Bigr )
-\tau \Bigl ({\bf n}, \bar {\bf n}, {\bf t},
\bar {\bf t}\Bigr )
\tau \Bigl ({\bf n}, \bar {\bf n}, {\bf t}^{[a_{\alpha}]},
\bar {\bf t}^{[b_{\alpha}]}\Bigr )
\\ &\\
&
=(ab)^{-1}\displaystyle{\left [ \vphantom{\frac{a}{b}}
\tau \Bigl ({\bf n}^{\alpha}, \bar {\bf n}_{\alpha}, 
{\bf t}^{[a_{\alpha}]},
\bar {\bf t}\Bigr )
\tau \Bigl ({\bf n}_{\alpha}, \bar {\bf n}^{\alpha}, {\bf t},
\bar {\bf t}^{[b_{\alpha}]}\Bigr ) -
\tau \Bigl ({\bf n}^{\alpha}, \bar {\bf n}^{\alpha}, 
{\bf t}^{[a_{\alpha}]},
\bar {\bf t}^{[b_{\alpha}]}\Bigr )
\tau \Bigl ({\bf n}_{\alpha}, \bar {\bf n}_{\alpha}, {\bf t},
\bar {\bf t}\Bigr ) \right ]},
\end{array}
\eeq

\beq\label{eq4even-2a}
\begin{array}{ll}
\bullet &
\tau \Bigl ({\bf n}, \bar {\bf n}^{\beta}, {\bf t}^{[a_{\alpha}]},
\bar {\bf t}\Bigr )
\tau \Bigl ({\bf n}, \bar {\bf n}^{\alpha}, {\bf t},
\bar {\bf t}^{[b_{\alpha}]}\Bigr )
-\tau \Bigl ({\bf n}, \bar {\bf n}^{\beta}, {\bf t},
\bar {\bf t}\Bigr )
\tau \Bigl ({\bf n}, \bar {\bf n}^{\alpha}, {\bf t}^{[a_{\alpha}]},
\bar {\bf t}^{[b_{\alpha}]}\Bigr )
\\ &\\
&\displaystyle{
=\epsilon_{\alpha \beta} a^{-1}\left [ \vphantom{\frac{a}{b}}
\tau \Bigl ({\bf n}^{\alpha}, \bar {\bf n}^{\alpha \beta}, 
{\bf t}^{[a_{\alpha}]},
\bar {\bf t}^{[b_{\alpha}]}\Bigr )
\tau \Bigl ({\bf n}_{\alpha}, \bar {\bf n}, {\bf t},
\bar {\bf t}\Bigr ) -
\tau \Bigl ({\bf n}^{\alpha}, \bar {\bf n}, 
{\bf t}^{[a_{\alpha}]},
\bar {\bf t}\Bigr )
\tau \Bigl ({\bf n}_{\alpha}, \bar {\bf n}^{\alpha \beta}, {\bf t},
\bar {\bf t}^{[b_{\alpha}]}\Bigr ) \right ], }
\end{array}
\eeq

\beq\label{eq4even-4a}
\begin{array}{ll}
\bullet &
\tau \Bigl ({\bf n}^{\beta}, \bar {\bf n}^{\beta}, {\bf t},
\bar {\bf t}\Bigr )
\tau \Bigl ({\bf n}^{\alpha}, \bar {\bf n}^{\alpha}, {\bf t}^{[a_{\alpha}]},
\bar {\bf t}^{[b_{\alpha}]}\Bigr )
-\tau \Bigl ({\bf n}^{\alpha \beta}, 
\bar {\bf n}^{\alpha \beta}, {\bf t}^{[a_{\alpha}]},
\bar {\bf t}^{[b_{\alpha}]}\Bigr )
\tau \Bigl ({\bf n}, \bar {\bf n}, {\bf t},
\bar {\bf t}\Bigr )
\\ &\\
&
=\tau \Bigl ({\bf n}^{\alpha}, \bar {\bf n}^{\beta}, 
{\bf t}^{[a_{\alpha}]},
\bar {\bf t}\Bigr )
\tau \Bigl ({\bf n}^{\beta}, \bar {\bf n}^{\alpha}, {\bf t},
\bar {\bf t}^{[b_{\alpha}]}\Bigr ) -
\tau \Bigl ({\bf n}^{\alpha \beta}, \bar {\bf n}, 
{\bf t}^{[a_{\alpha}]},
\bar {\bf t}\Bigr )
\tau \Bigl ({\bf n}, \bar {\bf n}^{\alpha \beta}, {\bf t},
\bar {\bf t}^{[b_{\alpha}]}\Bigr ).
\end{array}
\eeq

\beq\label{eq5even-4a}
\begin{array}{ll}
\bullet &
\tau \Bigl ({\bf n}, \bar {\bf n}^{\alpha}, {\bf t}^{[a_{\alpha}]},
\bar {\bf t}\Bigr )
\tau \Bigl ({\bf n}, \bar {\bf n}^{\beta}, {\bf t},
\bar {\bf t}^{[b_{\beta}]}\Bigr )
-\tau \Bigl ({\bf n}, \bar {\bf n}^{\alpha}, {\bf t},
\bar {\bf t}\Bigr )
\tau \Bigl ({\bf n}, \bar {\bf n}^{\beta}, {\bf t}^{[a_{\alpha}]},
\bar {\bf t}^{[b_{\alpha}]}\Bigr )
\\ &\\
&\displaystyle{
=\epsilon_{\alpha \beta} a^{-1}\left [ \vphantom{\frac{a}{b}}
\tau \Bigl ({\bf n}^{\alpha}, \bar {\bf n}, 
{\bf t}^{[a_{\alpha}]},
\bar {\bf t}\Bigr )
\tau \Bigl ({\bf n}_{\alpha}, \bar {\bf n}^{\alpha \beta}, {\bf t},
\bar {\bf t}^{[b_{\beta}]}\Bigr ) -
\tau \Bigl ({\bf n}^{\alpha}, \bar {\bf n}^{\alpha \beta}, 
{\bf t}^{[a_{\alpha}]},
\bar {\bf t}^{[b_{\beta}]}\Bigr )
\tau \Bigl ({\bf n}_{\alpha}, \bar {\bf n}, {\bf t},
\bar {\bf t}\Bigr ) \right ], }
\end{array}
\eeq

\beq\label{eq5even-7a}
\begin{array}{ll}
\bullet &
\tau \Bigl ({\bf n}^{\alpha}, \bar {\bf n}^{\alpha}, {\bf t}^{[a_{\alpha}]},
\bar {\bf t}\Bigr )
\tau \Bigl ({\bf n}^{\beta}, \bar {\bf n}^{\beta}, {\bf t},
\bar {\bf t}^{[b_{\beta}]}\Bigr )
-\tau \Bigl ({\bf n}^{\beta}, 
\bar {\bf n}^{\alpha}, {\bf t},
\bar {\bf t}\Bigr )
\tau \Bigl ({\bf n}^{\alpha}, \bar {\bf n}^{\beta}, {\bf t}^{[a_{\alpha}]},
\bar {\bf t}^{[b_{\beta}]}\Bigr )
\\ &\\
&
=\tau \Bigl ({\bf n}^{\alpha \beta}, \bar {\bf n}^{\alpha \beta}, 
{\bf t}^{[a_{\alpha}]},
\bar {\bf t}^{[b_{\beta}]}\Bigr )
\tau \Bigl ({\bf n}, \bar {\bf n}, {\bf t},
\bar {\bf t}\Bigr ) -
\tau \Bigl ({\bf n}^{\alpha \beta}, \bar {\bf n}, 
{\bf t}^{[a_{\alpha}]},
\bar {\bf t}\Bigr )
\tau \Bigl ({\bf n}, \bar {\bf n}^{\alpha \beta}, {\bf t},
\bar {\bf t}^{[b_{\beta}]}\Bigr ).
\end{array}
\eeq

\beq\label{eq5even-3a}
\begin{array}{ll}
\bullet &
\tau \Bigl ({\bf n}, \bar {\bf n}, {\bf t}^{[a_{\alpha}]},
\bar {\bf t}\Bigr )
\tau \Bigl ({\bf n}, \bar {\bf n}, {\bf t},
\bar {\bf t}^{[b_{\beta}]}\Bigr )
-\tau \Bigl ({\bf n}, \bar {\bf n}, {\bf t},
\bar {\bf t}\Bigr )
\tau \Bigl ({\bf n}, \bar {\bf n}, {\bf t}^{[a_{\alpha}]},
\bar {\bf t}^{[b_{\beta}]}\Bigr )
\\ &\\
&
=(ab)^{-1}\displaystyle{\left [ \vphantom{\frac{a}{b}}
\tau \Bigl ({\bf n}^{\alpha}, \bar {\bf n}_{\beta}, 
{\bf t}^{[a_{\alpha}]},
\bar {\bf t}\Bigr )
\tau \Bigl ({\bf n}_{\alpha}, \bar {\bf n}^{\beta}, {\bf t},
\bar {\bf t}^{[b_{\beta}]}\Bigr ) -
\tau \Bigl ({\bf n}^{\alpha}, \bar {\bf n}^{\beta}, 
{\bf t}^{[a_{\alpha}]},
\bar {\bf t}^{[b_{\beta}]}\Bigr )
\tau \Bigl ({\bf n}_{\alpha}, \bar {\bf n}_{\beta}, {\bf t},
\bar {\bf t}\Bigr ) \right ]},
\end{array}
\eeq

\beq\label{eq2even-2a}
\begin{array}{ll}
\bullet &
\tau \Bigl ({\bf n}, \bar {\bf n}^{\alpha }, {\bf t}^{[a_{\alpha}]},
\bar {\bf t}\Bigr )
\tau \Bigl ({\bf n}, \bar {\bf n}^{\beta }, {\bf t}^{[b_{\alpha}]},
\bar {\bf t}\Bigr )
-\tau \Bigl ({\bf n}, \bar {\bf n}^{\alpha}, {\bf t}^{[b_{\alpha}]},
\bar {\bf t}\Bigr )
\tau \Bigl ({\bf n}, \bar {\bf n}^{\beta }, {\bf t}^{[a_{\alpha}]},
\bar {\bf t}\Bigr )
\\ &\\
&\displaystyle{
=\epsilon_{\alpha \beta}(a^{-1}-b^{-1})\left [ \vphantom{\frac{a}{b}}
\tau \Bigl ({\bf n}^{\alpha}, \bar {\bf n}, 
{\bf t}^{[a_{\alpha}b_{\alpha}]},
\bar {\bf t}\Bigr )
\tau \Bigl ({\bf n}_{\alpha}, \bar {\bf n}^{\alpha \beta}, {\bf t},
\bar {\bf t}\Bigr ) \right. }
\\ &\\
&\displaystyle{\phantom{aaaaaaaaaaaaaaaaaaaaaa}
-\left. \vphantom{\frac{a}{b}}
\tau \Bigl ({\bf n}^{\alpha}, \bar {\bf n}^{\alpha \beta}, 
{\bf t}^{[a_{\alpha}b_{\alpha}]},
\bar {\bf t}\Bigr )
\tau \Bigl ({\bf n}_{\alpha}, \bar {\bf n}, {\bf t},
\bar {\bf t}\Bigr ) \right ], }
\end{array}
\eeq

\beq\label{eq3even-2a}
\begin{array}{ll}
\bullet &
\tau \Bigl ({\bf n}^{\alpha}, \bar {\bf n}^{\alpha}, {\bf t}^{[a_{\alpha}]},
\bar {\bf t}\Bigr )
\tau \Bigl ({\bf n}^{\beta}, \bar {\bf n}^{\beta}, {\bf t}^{[b_{\beta}]},
\bar {\bf t}\Bigr )
-\tau \Bigl ({\bf n}^{\beta}, 
\bar {\bf n}^{\alpha}, {\bf t}^{[b_{\beta}]},
\bar {\bf t}\Bigr )
\tau \Bigl ({\bf n}^{\alpha}, \bar {\bf n}^{\beta}, {\bf t}^{[a_{\alpha}]},
\bar {\bf t}\Bigr )
\\ &\\
&
=\tau \Bigl ({\bf n}^{\alpha \beta}, \bar {\bf n}^{\alpha \beta}, 
{\bf t}^{[a_{\alpha}b_{\beta}]},
\bar {\bf t}\Bigr )
\tau \Bigl ({\bf n}, \bar {\bf n}, {\bf t},
\bar {\bf t}\Bigr ) -
\tau \Bigl ({\bf n}^{\alpha \beta}, \bar {\bf n}, 
{\bf t}^{[a_{\alpha}b_{\beta}]},
\bar {\bf t}\Bigr )
\tau \Bigl ({\bf n}, \bar {\bf n}^{\alpha \beta}, {\bf t},
\bar {\bf t}\Bigr ).
\end{array}
\eeq

\section*{Acknowledgments}
\addcontentsline{toc}{section}{Acknowledgments}

This work is an output of the research project 
``Symmetry. Information. Chaos''
implemented as a part of the Basic Research Program at 
National Research University Higher School 
of Economics (HSE University).


\begin{thebibliography}{99}

\addcontentsline{toc}{section}{References}

\bibitem{DJKM83} E. Date, M. Jimbo, M. Kashiwara and T. Miwa, {\it Transformation groups
for soliton equations}, in: M. Jimbo, T. Miwa (Eds.), Nonlinear Integrable Systems --
Classical and Quantum, World Scientific, 1983, pp. 39--120.

\bibitem{JM83} M. Jimbo and T. Miwa, {\it Solitons and 
infinite dimensional Lie
algebras}, Publ. Res. Inst. Math. Sci. Kyoto {\bf 19} (1983) 943--1001.




\bibitem{HO} R. Hirota and Y. Ohta, {\it Hierarchies of coupled 
soliton equations I}, J. Phys. Soc. Japan {\bf 60} (1991) 798-809.

\bibitem{AHM} M. Adler, E. Horozov and P. van Moerbeke, {\it 
The Pfaff lattice and skew-orthogonal polynomials}, Int. Math. Res.
Notices {\bf 1999} (1999), no 11, 569--588.

\bibitem{ASM} M. Adler, T. Shiota and P. van Moerbeke, {\it
Pfaff $\tau$-functions}, Math. Ann. {\bf 322} (2002) 423--476.


\bibitem{Kakei} S. Kakei, {\it Orthogonal and symplectic matrix integrals
and coupled KP hierarchy}, J. Phys. Soc. Japan {\bf 99} (1999) 2875--2877.



\bibitem{IWS} S. Isojima, R. Willox and J. Satsuma, {\it On various 
solutions of the coupled KP equation}, J. Phys. A: Math. Gen.
{\bf 35} (2002) 6893--6909.

\bibitem{Willox} R. Willox, {\it On a generalized Tzitzeica 
equation}, Glasgow Math. J. {\bf 47A} (2005) 221--231.




\bibitem{Kodama} Y. Kodama and K.-I. Maruno, {\it $N$-soliton 
solutions to the DKP hierarchy and the Weyl group actions}, 
J. Phys. A: Math. Gen. {\bf 39} (2006) 4063--4086.



\bibitem{Takasaki07} K. Takasaki, {\it 
Differential Fay identities and auxiliary linear 
problem of integrable hiearchies},  
Advanced Studies in Pure Mathematics {\bf 61} (2011) 387--441.

\bibitem{Takasaki09} K. Takasaki, {\it Auxiliary linear problem,
difference Fay identities and dispersionless limit of Pfaff-Toda hierarchy},
SIGMA {\bf 5} (2009) 109.



\bibitem{DJKM81} E. Date, M. Jimbo, M. Kashiwara and 
T. Miwa, {\it Transformation groups
for soliton equations III}, J. Phys. Soc. Japan {\bf 50} (1981) 3806--3812.

\bibitem{KL93} V. Kac and J. van de Leur, {\it The $n$-component KP hierarchy and 
representation theory}, in: A.S. Fokas, V.E. Zakharov (Eds.), 
Important Developments
in Soliton Theory, Springer-Verlag, Berlin, Heidelberg, 1993.



\bibitem{Teo11} L.-P. Teo, {\it The multicomponent KP hierarchy: 
differential Fay identities and Lax
equations}, J. Phys. A: Math. Theor. {\bf 44} (2011) 225201.

\bibitem{TT07} K. Takasaki and T. Takebe, {\it Universal Whithem hierarchy,
dispersionless Hirota equations and multicomponent KP hierarchy}, 
Physica D {\bf 235} (2007) 109--125.

\bibitem{SZ24}
A. Savchenko, A. Zabrodin, {\it Multicomponent DKP 
hierarchy and its dispersionless limit},
Lett. Math. Phys. {\bf 115:22} (2025).


\bibitem{Miwa82}
T. Miwa, {\it On Hirota's difference equations}, Proc. Japan Acad. 
{\bf 58} Ser. A (1982) 9--12.

\bibitem{Shigyo13}
Y. Shigyo, {\it On addition formulae of KP, mKP
and BKP hierarchies}, SIGMA {\bf 9} (2013) 035.

\bibitem{TT95}
K. Takasaki and T. Takebe, {\it Integrable hierarchies and dispersionless
limit}, Rev. Math. Phys. {\bf 7} (1995) 743--808. 




\bibitem{AZ14} V. Akhmedova and A. Zabrodin, {\it Dispersionless
DKP hierarchy and elliptic L\"owner equation}, J. Phys. A: Math. Theor.
{\bf 47} (2014) 392001, arXiv:1404.5135.

\bibitem{Z24}
A. Zabrodin, {\it Dispersionless version of 
the multicomponent KP hierarchy revisited},
Physica D {\bf 467} (2024) 134286,
arXiv:2404.10406. 



\bibitem{TZ25}
T. Takebe, A. Zabrodin, {\it Multi-component Toda lattice hierarchy},
Russian Mathematical Surveys, {\bf 80:4} (2025)
591--665.




\end{thebibliography}
\end{document}